%% file: sphere-dataset.tex
\documentclass[11pt]{article}
\pdfoutput=1
\usepackage{amsmath}
\usepackage{graphicx}
\usepackage{url}
\usepackage{subfig}
\usepackage{xspace}
\usepackage{hyperref}
\usepackage{booktabs}
\usepackage{xspace}
\usepackage{xcolor}
\usepackage{listings}
\usepackage[labelfont=bf]{caption}
\usepackage{textcomp} 
\usepackage{authblk}

\newcommand{\ie}{{\it i.e.},~}
\newcommand{\eg}{{\it e.g.},~}

\newcommand\code[1]{\texttt{#1}}

\newcommand{\degc}{$^o$C\xspace}

\newcommand{\linesep}{\noindent\rule{\textwidth}{1pt}\xspace}

\lstdefinelanguage{json}{
    basicstyle=\scriptsize\ttfamily,
    stepnumber=1,
    numbersep=8pt,
    showstringspaces=false,
    breaklines=true,
    frame=lines,
    backgroundcolor=\color{background},
    literate=
     *{:}{{{\color{punct}{:}}}}{1}
     {,}{{{\color{punct}{,}}}}{1}
     {'}{{{\color{punct}{'}}}}{1}
     {\{}{{{\color{delim}{\{}}}}{1}
     {\}}{{{\color{delim}{\}}}}}{1}
     {[}{{{\color{delim}{[}}}}{1}
     {]}{{{\color{delim}{]}}}}{1},
}

\definecolor{dkgreen}{rgb}{0,0.6,0}
\definecolor{dkblue}{rgb}{0,0,0.6}
\definecolor{dkred}{rgb}{0.6,0,0}
\definecolor{grey}{rgb}{0.5,0.5,0.5}

\colorlet{punct}{red!60!black}
\definecolor{background}{HTML}{EEEEEE}
\definecolor{delim}{RGB}{20,105,176}
\colorlet{numb}{magenta!60!black}

\begin{document}

\title{A Guide to the SPHERE 100 Homes Study Dataset}

\author[1]{Atis~Elsts}
\author[1]{Tilo Burghardt}
\author[1]{Dallan~Byrne}
\author[1]{Massimo~Camplani}
\author[1]{Dima Damen}
\author[1,3]{Xenofon~Fafoutis}
\author[1]{Sion~Hannuna}
\author[2]{William Harwin}
\author[1]{Michael~Holmes}
\author[2]{Balazs Janko}
\author[1]{V\'{i}ctor~Ponce-L\'{o}pez}
\author[1]{Alessandro~Masullo}
\author[1]{Majid Mirmehdi}
\author[1]{George~Oikonomou}
\author[1]{Robert~Piechocki}
\author[2]{R.~Simon~Sherratt}
\author[1]{Emma~Tonkin}
\author[1]{Niall~Twomey}
\author[1]{Antonis~Vafeas}
\author[1]{Przemyslaw~Woznowski}
\author[1]{Ian~Craddock}

\affil[1]{University of Bristol, Faculty of Engineering}
\affil[2]{University of Reading, Department of Biomedical Engineering}
\affil[3]{Technical University of Denmark}
\setcounter{Maxaffil}{0}
\renewcommand\Affilfont{\itshape\small}

\date{%
  Technical Report \\
  \today
}

\maketitle

\begin{abstract}
 The SPHERE project has developed a multi-modal sensor platform for health and behavior monitoring in residential environments. So far, the SPHERE platform has been deployed for data collection in approximately 50 homes for duration up to one year.
This technical document describes the format and the expected content of the SPHERE dataset(s) under preparation.
It includes a list of some data quality problems (both known to exist in the dataset(s) and potential ones), their workarounds, and other information important to people working with the SPHERE data, software, and hardware.
This document does not aim to be an exhaustive descriptor of the SPHERE dataset(s); it also does not aim to discuss or validate the potential scientific uses of the SPHERE data.

\end{abstract}

\tableofcontents

\section{Introduction}

SPHERE (a Sensing Platform for HEalthcare in a Residential Environment) is a multipurpose, multi-modal platform of non-medical home sensors that aims to serve as a prototype for future residential healthcare systems~\cite{zhu2015bridging}.
The SPHERE system has been installed in approximately 50 homes in Bristol, UK, in each of which it collected data for duration up to one year.
The data from these installations will be consolidated in one or more datasets after the end of the project, and made available to researchers where ethical considerations and policies permit.

Different aspects of the SPHERE system have been already been described in research papers~\cite{woznowski2015multi, Elsts2017Lessons,Fafoutis2018Lessons}.
However, none of these papers serves as a reference to the data produced by the SPHERE system. This technical report aims to close this gap. 
To clarify, this work is not a data descriptor for the 100 Homes Study: it is a technical reference manual that may be used to support working with the SPHERE system and with datasets generated through the SPHERE system, including the data collected from the SPHERE 100 Homes Study.

The SPHERE system produces multiple modalities of data (Section~\ref{sec:df}), namely: environmental sensor data, wearable data, and video data.
Most of the SPHERE devices also produce monitoring data, such as the data about the health status of the devices, their network connections and any problems observed.
The SPHERE platform includes on hardware designed specifically for the project: for example, the wearable sensors~\cite{fafoutis2017designing}, the environmental sensors~SPES \cite{SPES-2}, and the IoT (Internet of Things) gateways. In contrast, the video subsystems uses commercial-off-the-shelf hardware~\cite{hall2016designing}.
Many parts of the SPHERE software stack either already are open source or may be released as open source in the future.
The SPHERE technologies are already have been successfully used in related projects, such as in EurValve~\cite{pope2017sphere}, to monitor of patients before and after a heart-valve surgery in their homes, and in HemiSPHERE \cite{holmes_analysis_2018}, to monitor patients that have undergone surgeries for hip or knee replacement.

The main goals of this document are to: (1) serve as a reference of the SPHERE dataset and (2) facilitate future applications of  SPHERE software and hardware.

\section{Executive summary}

\begin{figure}
  \centering
  \includegraphics[width=0.4\linewidth]{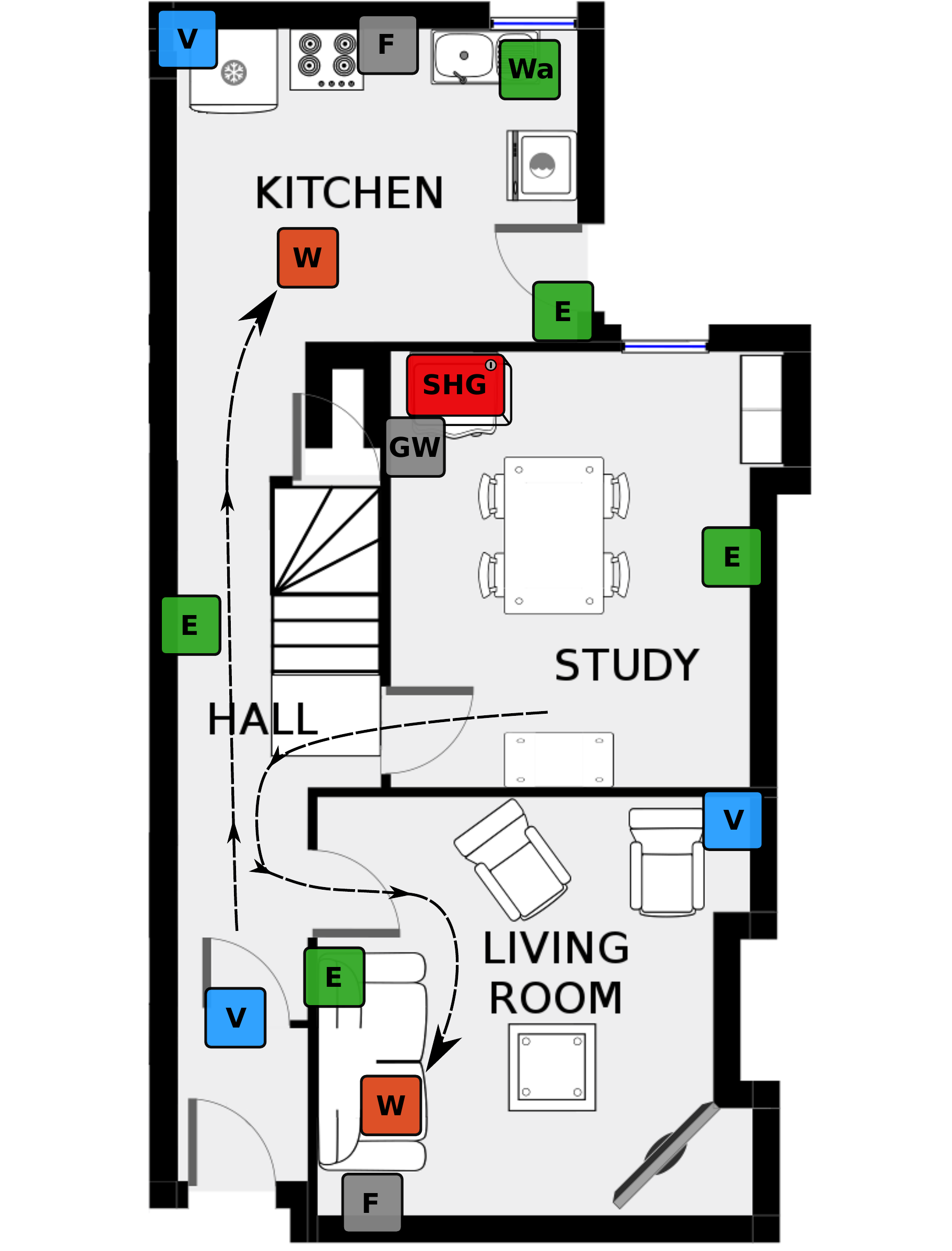}
  \vspace{-4pt}
  \caption{\textbf{SPHERE device layout in the first floor of a residential house.}
    $SHG$ -- the SPHERE Home Gateway.
    $W$ -- mobile wearables.
    $V$ -- video cameras with the attached video gateways.
    $E$ -- environmental sensors.
    $Wa$ -- the water sensor.
    $F$ -- forwarding gateways.
    $GW$ -- the main forwarding gateway attached to the SHG with a cable.
}
  \label{fig:sphere-house}
\end{figure}

The SPHERE platform consists of the following devices that produce data:
\begin{itemize}
\item \textbf{Wearable nodes} (``Wristband sensors'', Fig.~\ref{fig:spw}). These are wrist-worn accelerometer sensors. There is one wearable per each inhabitant. They are worn continuously, including during sleep. These help with indoor localization and with activity monitoring.
\item \textbf{Video cameras} (``Silhouette sensors'', Fig.~\ref{fig:videocamera}). There are up to three silhouette sensors per house, deployed in the hall, living room, and kitchen. These sensors help to monitor the quality of movement of the participants. These cameras do not record sound.
\item \textbf{Video gateways} (Fig.~\ref{fig:videonuc}). These are small computers that process the raw data from the video cameras and automatically extract the silhouettes of the persons in the field of the view. Only the silhouettes and the ``boxes'' around them are stored in the dataset, not the raw video.
\item \textbf{Environmental sensors} (Fig.~\ref{fig:spes}). These nodes monitor light, temperature, air humidity, and barometric pressure levels. They also have a room presence sensor, as used in security alarms, so they can detect people moving in their field of view. There is one environmental sensor per each room.
\item \textbf{Water flow sensors} (Fig.~\ref{fig:water}). This is a modified version of the environmental sensor. It reports the same values as the environmental sensors, as well as the water flow status of the cold and hot water taps (``on''/''off'' for each of them).
  Only one such sensor per house is deployed, located in the kitchen, under the sink.
\item \textbf{Forwarding gateways} (``Receivers'', Fig.~\ref{fig:spg2}). The main functions of these devices are to forward the data from the wearable nodes and environmental sensors to the SPHERE Home gateway, as well as assist in the indoor localization by continuously measuring the radio signal levels from the wearable nodes. There is one sensor per approximately each two rooms. They also monitor a few environmental parameters.
\item \textbf{Appliance monitors} (``CurrentCost sensors'', Fig.~\ref{fig:ccs}). These sensors are plugged in some wall sockets and record the electricity usage of the attached appliances. These appliances  are typically equipped with sensors: kettle, microwave oven, toaster, TV.
\item \textbf{The SPHERE Genie.} This is a tablet computer that allows the user to control the SPHERE system, such as to pause the collection of the data. It produces command and notification messages.
\item \textbf{A SPHERE Home Gateway} (Fig.~\ref{fig:shg}). This is a central server that wirelessly collects the data from the rest of the SPHERE system and stores it in a local, encrypted hard disk. It produces no sensor data on its own, but does generate some monitoring data.
\end{itemize}

\section{Architecture}

In the \emph{SPHERE 100 Homes Study}, the SPHERE platform is deployed in residential houses.
In such a house (see Fig.~\ref{fig:sphere-house} for an example), all SPHERE devices are connected in a network, or, more accurately, multiple networks, including WiFi, IEEE 802.15.4, and BLE (Bluetooth Low Energy) networks: see~\cite{woznowski2015multi, Elsts2017Lessons,hall2016designing} for the details.
All data produced by the SPHERE devices is eventually delivered to and stored on the SPHERE Home Gateway (SHG).
On the SHG, the data is temporarily stored in a Mongo database. Once per day, the contents of this database are saved as \code{.bson} files; these files are then stored on an external hard drive attached to the SHG.

SPHERE uses MQTT (Message Queuing Telemetry Transport) protocol as its backbone publish/subscribe system. The subsystems of SPHERE deliver their data to the database through ActiveMQ MQTT broker that is running on the SPHERE Home Gateway.

The main data and monitoring data sources in SPHERE are the wearables, the environmental sensors, and the video cameras/gateways.
Data from both wearables and environmental sensors are delivered to the MQTT broker through a gateway script, which takes the sensor data items, puts them in timestamped JSON documents, and sends the documents to the MQTT broker.
The video gateway data does not go through this gateway script: they produce MQTT messages themselves.
Other data sources include the appliance electricity consumption sensors. 
Finally, scripts running on the SPHERE Home Gateway produce system monitoring data.

%



\input{dataformats.tex}

\input{dataquality.tex}

\section{Conclusion}
In this report we describe the data format of the SPHERE dataset, as well as some known data quality problems and limitations.
Our hope is that this document will be useful to those who are looking to analyze and extract value from the SPHERE dataset, as well as to projects that plan to use SPHERE software or hardware.

\section*{Acknowledgments}

The SPHERE Interdisciplinary Research Collaboration (IRC) includes 50+ researchers and support personnel working together and contributing to the SPHERE vision.
The authors acknowledge the valued contribution, discussions and exchange of ideas that have influenced the SPHERE platform and shaped the SPHERE dataset.
A full list of SPHERE collaborators can be found at http://www.irc-sphere.ac.uk.

This work was performed under the SPHERE IRC funded by the UK Engineering and Physical Sciences Research Council (EPSRC), Grant EP/K031910/1. NT is currently supported by "Continuous Behavioural Biomarkers of Cognitive Impairment" project funded by the UK Medical Research Council Momentum Awards under Grant MC/PC/16029.

\bibliographystyle{IEEEtran}
\bibliography{IEEEabrv,sphere-dataset.bib}

\end{document}

%% file: dataformats.tex
\section{Data format}
\label{sec:df}

SPHERE stores data in JSON (JavaScript Object Notation) documents. The data format in the SPHERE dataset uses SenML (Sensor Markup Language) \cite{senml}, extended with extra attributes where needed.

\subsection{General document format}

These fields are present at the top level of SPHERE documents:

\begin{itemize}
\item \code{hid} -- house identifier.
\item \code{uid} -- identifier of the component that created this document. Could be an IPv6 address, a MAC address, or a name.
\item \code{bt} -- base time: the time this sensor record was created. A UTC timestamp (in some cases added by the SHG when the data packet is received) compliant with ISO 8601 data elements and interchange formats. 
\item \code{\_id} -- MongoDB record identifier. Could be used to infer the time when this sensor record was added to the database. Usually this time value is slightly later than \code{bt} (up to two minutes for some documents); it should never be earlier than \code{bt}.
\item \code{ts} -- timestamp: for details see the note below.
\item \code{tso} -- timestamp offset: for details see the note below.
\item \code{mc} -- ``message count'', i.e., sequence number. Note that for some sequence numbers wraparounds can be observed in SPHERE data: for example, wearable packets have 16-bit sequence number, which wraps around after every 65536 packets.
\item \code{e} --  an array containing sensor data.
\item \code{gw} -- for wearable messages, an array containing data added by the gateways.
\item \code{type} -- document type, e.g. \code{ALERT}. Not present for sensor data documents.
\end{itemize}

\subsubsection{Timestamps}

There are multiple timestamp-related fields in many documents, for example, \code{ts} and \code{tso}.
With these two fields in particular, there are two separate cases:
\begin{enumerate}
\item \code{tso} is zero. In this case, \code{ts} is simply the UNIX timestamp in seconds.
\item \code{tso} is nonzero. In this case, \code{ts} is expressed in TSCH network ticks (using $\approx 1/100$ of a second as the units), and the \code{tso} is the UNIX timestamp (in seconds) of the moment when the TSCH network's time counting was started.
\end{enumerate}
In both cases, the sum of \code{ts} and \code{tso} fields, if both are present, should be always equal to the UNIX timestamp corresponding to the \code{bt} field.

When working with the SPHERE data, one normally does not need to use \code{ts} and \code{tso} field directly, but can use the \code{bt} field instead. The other two fields are there only for timestamp integrity checking and potential restoration purposes; it introduces some redundancy in the data and serves as an extra safety measure.

\subsection{Which devices produce which data?}

Note that not every message from a device contains all sensors.

\begin{itemize}
\item SPW-2 wearable sensor nodes (Fig.~\ref{fig:spw}, \code{SPW2} in the monitoring data)
  \begin{itemize}
  \item[+]     \code{ACCEL}
  \end{itemize}
  
  Note that there's also an array named \code{gw} in the documents with  \code{ACCEL} data. This array contains the RSSI information.

\begin{figure}[ht!]
\centering
  \includegraphics[width=0.5\linewidth]{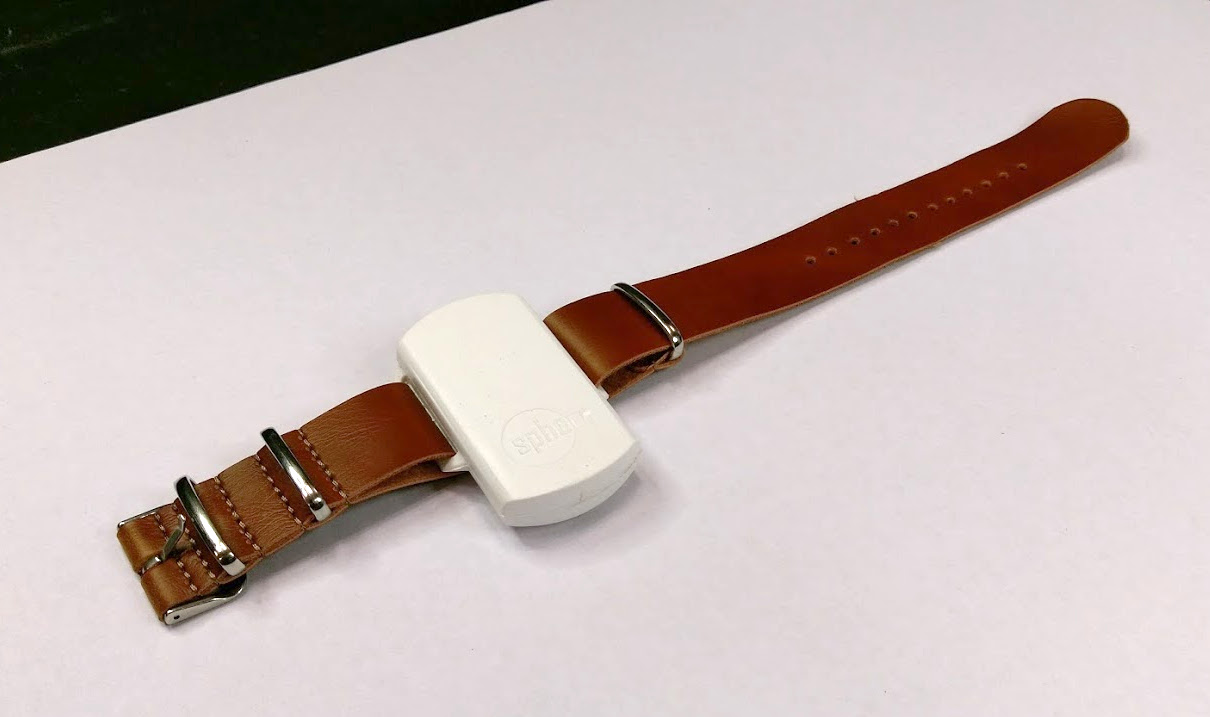}
  \caption{SPHERE Wearable node.}
  \label{fig:spw}
\end{figure}

\item Video NUC (Fig.~\ref{fig:videonuc}) with associated 3D camera (Fig.~\ref{fig:videocamera})
  \begin{itemize}
  \item[+]     RSSI data about messages from wearables.
  \item[+]     Video data (silhouettes and bounding boxes)
  \end{itemize}
  
Note that raw video data is not retained: only visual features computed and extracted from them are retained and stored.
 
\begin{figure}[ht!]
\centering
  \includegraphics[width=0.3\linewidth]{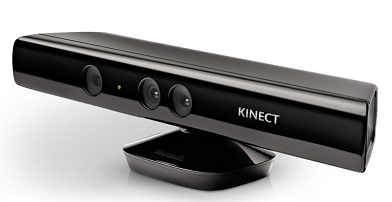}
  \caption{SPHERE Video Camera.}
  \label{fig:videocamera}
\end{figure}

\begin{figure}[ht!]
\centering
  \includegraphics[width=0.3\linewidth]{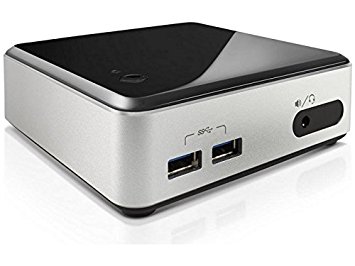}
  \caption{SPHERE Video NUC.}
  \label{fig:videonuc}
\end{figure}

\item SPES-2 environmental sensor nodes (Fig.~\ref{fig:spes}, \code{SPES2} in the monitoring data)
  \begin{itemize}
  \item[+]     \code{HDC\_TEMP}
  \item[+]     \code{HDC\_HUM}
  \item[+]     \code{BMP\_TEMP}
  \item[+]     \code{BMP\_PRES}
  \item[+]     \code{LT}
  \item[+]     \code{PIR\_TRIGS}
  \end{itemize}

\begin{figure}[ht!]
\centering
  \includegraphics[width=0.3\linewidth]{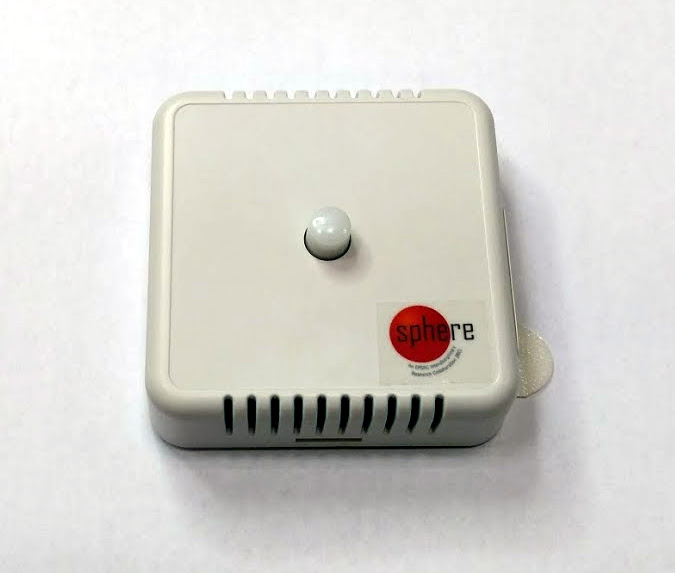}
  \caption{SPHERE Environmental Sensor node.}
  \label{fig:spes}
\end{figure}

\item SPES-2 \textbf{water flow} sensor nodes (Fig.~\ref{fig:water}, \code{SPAQ2\_P} in the monitoring data -- ``AQ'' standing for ``Aqua'', ``P'' standing for ``piezo'')
  \begin{itemize}
  \item[+]     \code{HDC\_TEMP}
  \item[+]     \code{HDC\_HUM}
  \item[+]     \code{BMP\_TEMP}
  \item[+]     \code{BMP\_PRES}
  \item[+]     \code{LT}
  \item[+]     \code{PIR\_TRIGS}
  \item[+]     \code{PIEZO\_WATER\_C}
  \item[+]     \code{PIEZO\_WATER\_H}
  \end{itemize}
   
  Based on the raw sensor values \code{PIEZO\_WATER\_C} (cold) and \code{PIEZO\_WATER\_H} (hot) the gateway script publishes estimated water flow status in \code{FLOW\_COLD\_ON} and \code{FLOW\_HOT\_ON}.

\begin{figure}[ht!]
\centering
  \includegraphics[width=0.6\linewidth]{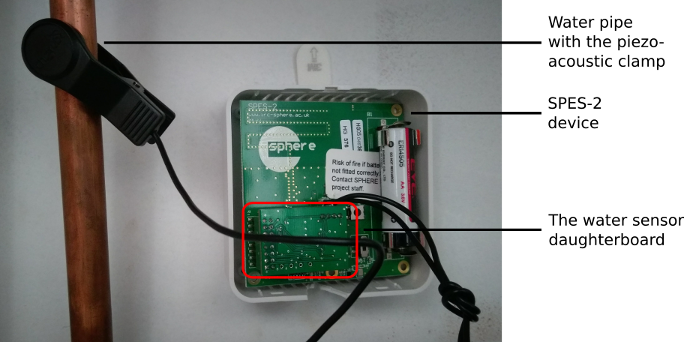}
  \caption{SPHERE Environmental Water Sensor.}
  \label{fig:water}
\end{figure}

\item SPG-2 forwarding gateway nodes (Fig.~\ref{fig:spg2}, \code{SPG2\_F} in the monitoring data):

  \begin{itemize}
  \item[+] \code{HDC\_TEMP}
  \item[+] \code{HDC\_HUM}
  \item[+] \code{BMP\_TEMP}
  \item[+] \code{BMP\_PRES}
  \item[+] \code{WEARABLE\_ADV} -- internal only, contains wearable acceleration data
  \end{itemize}

\begin{figure}[ht!]
\centering
  \includegraphics[width=0.4\linewidth]{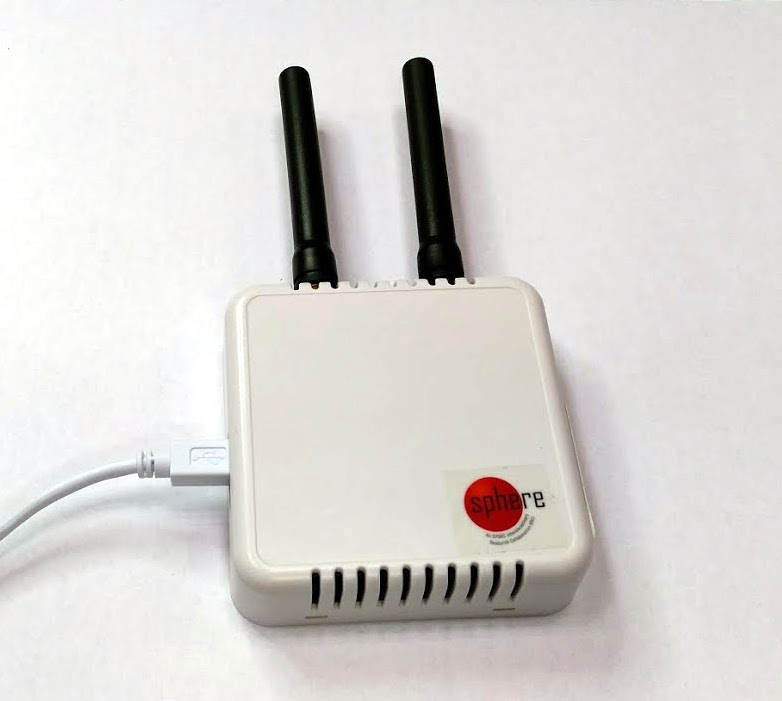}
  \caption{SPHERE Forwarding Gateway node.}
  \label{fig:spg2}
\end{figure}

\item Appliance monitor sensors (Fig.~\ref{fig:ccs})
  \begin{itemize}
  \item[+]     \code{ELEC}
  \end{itemize}
  
\begin{figure}[ht!]
\centering
  \includegraphics[width=0.4\linewidth]{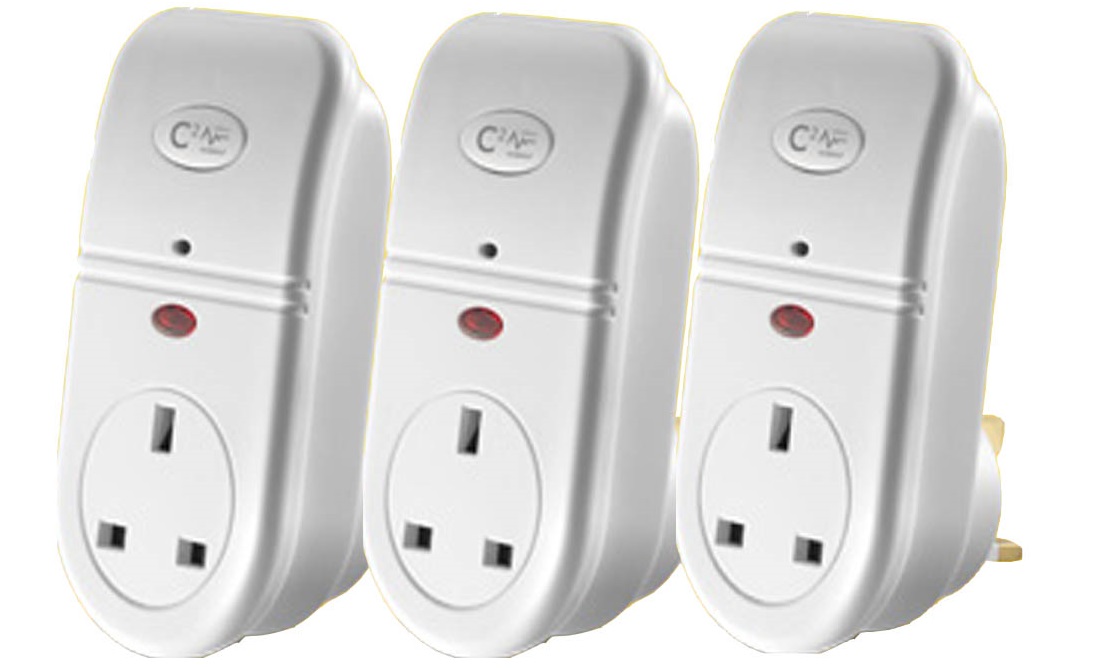}
  \caption{Appliance monitor (\emph{CurrentCost}) sensor nodes.}
  \label{fig:ccs}
\end{figure}

\item SPG-2 border router nodes (Fig.~\ref{fig:spg2br}, \code{SPG2\_BR} in the monitoring data): no sensors, monitoring information only.

  \begin{figure}[ht!]
\centering
  \includegraphics[width=0.5\linewidth]{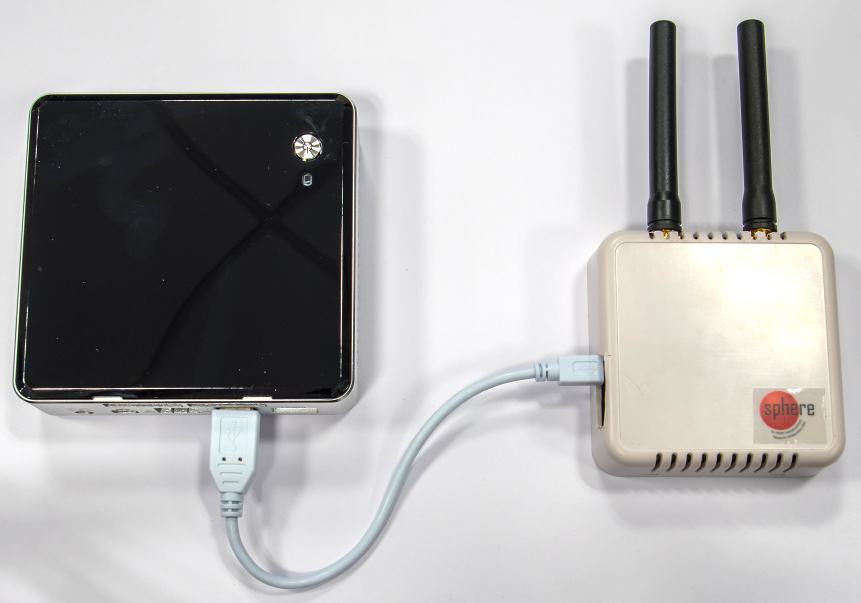}
  \caption{SPHERE Home Gateway with a connected SPG-2 node.}
  \label{fig:spg2br}
  \label{fig:shg}
\end{figure}

\item Home Gateway NUC (Fig.~\ref{fig:shg})
  \begin{itemize}
  \item[+]     RSSI data about messages from wearables.
  \item[+]     MQTT monitoring data.
  \item[+]     Nagios monitoring data.
  \end{itemize}

\end{itemize}

\subsection{Environmental sensor data}

Environmental sensor document example:

\begin{lstlisting}[language=json]
{
	'_id': ObjectId('5a005121b190070a60eed46f'),
	'e': [{
		'n': 'PIR_TRIGS',
		'v': [0, 0, 0, 0, 0, 0, 0, 0, 0, 0, 0, 0, 0, 0, 0, 0, 0, 0, 0, 0, 0, 0, 0, 0, 0, 0, 0, 0, 0, 0]
	}],
	'bt': datetime.datetime(2017, 11, 6, 12, 9, 56, 950000),
	'ts': 80569776,
	'uid': 'fd00::212:4b00:0:ff80',
	'tso': 1509164499.1906219,
	'mc': 94855,
	'hid': '0000'
}
\end{lstlisting}

\subsubsection{Temperature, humidity, pressure and light data}

  \begin{itemize}
  \item     \code{HDC\_TEMP} -- HDC sensor's temperature value, in \degc.
  \item     \code{HDC\_HUM} -- HDC sensor's relative humidity value, in \%.
  \item     \code{BMP\_TEMP} -- BMP sensor's temperature value, in \degc.
  \item     \code{BMP\_PRES} -- BMP sensor's air pressure  value, in hPa.
  \item     \code{LT} -- light sensor's value, in luxes. Note that the light sensors are placed \emph{in} the enclosures, therefore were low light readings are normally expected.
  \end{itemize}

\subsubsection{PIR sensor data}

\begin{figure}[h]
\centering
  \includegraphics[width=0.7\linewidth]{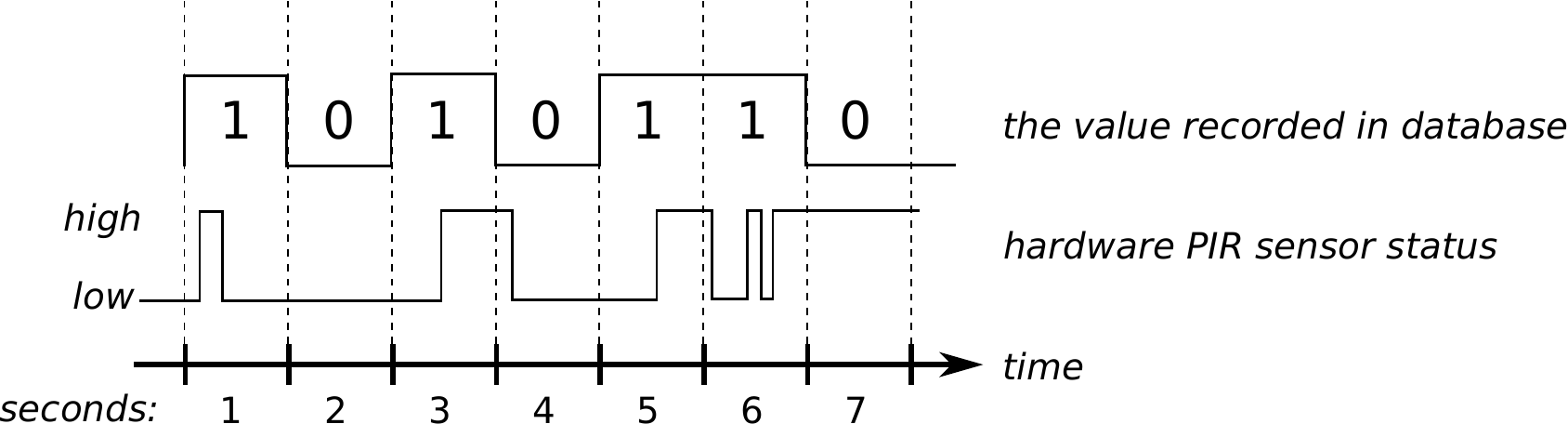}
  \caption{Example of PIR sensor triggers and the corresponding data recorded in the database.}
  \label{fig:pir}
\end{figure}

  \begin{itemize}
  \item     \code{PIR\_TRIGS} - PIR sensor's trigger history. 
  \end{itemize}
  \code{PIR\_TRIGS} is an array with 30 samples, describing the PIR status during the last 30 seconds. The timestamp of the document refers to the moment the last sample was collected, i.e. these 30 samples correspond to the 30 seconds immediately before the timestamp of the document. Each sample is either 1 or 0: it is 1 for those seconds when the PIR sensor was triggered during any moment in that second (seconds 1, 3, 5, and 6 in Fig.~\ref{fig:pir}), and 0 otherwise.

  Note that the \texttt{PIR\_TRIGS} messages are generated \textit{every 20 seconds}, not 30 seconds, as would be intuitive.
  This means there is some duplication in the PIR data as the first 10 entries of the PIR data message are duplicates of the last 10 entries of the previous PIR data message.
  This is an accidental behavior due to a typo in the program code.

  The values stored in the database effectively correspond to a rising edge detector that takes the ``raw'' (hardware) PIR status as the input.
  Figure \ref{fig:pir} gives an example of how the physical PIR sensor triggers correspond to the values recorded in the database. Here are some examples of how it works:
  \begin{itemize}
  \item The sensor goes up, then down again: the value in the database is 1 (the 1st second in Fig.~\ref{fig:pir}).
  \item The sensor goes up and not down: the value in the database is 1 (the 3rd second in Fig.~\ref{fig:pir}).
  \item The sensor goes down and not up: the value in the database is 0 (the 4th second in Fig.~\ref{fig:pir}).
  \item The sensor goes up/down multiple times: the value in the database is 1 (the 6th second in Fig.~\ref{fig:pir}).
  \item The value of the hardware PIR sensor is \emph{low} during the whole second: the value in the database is 0 (the 2nd second in Fig.~\ref{fig:pir}).
  \item The value of the hardware PIR sensor is \emph{high} during the whole second: the value in the database is 0 (the 7th second in Fig.~\ref{fig:pir}).
  \end{itemize}

\subsubsection{Water flow sensor data}

 \begin{figure}[ht!]
\centering
  \includegraphics[width=0.4\linewidth]{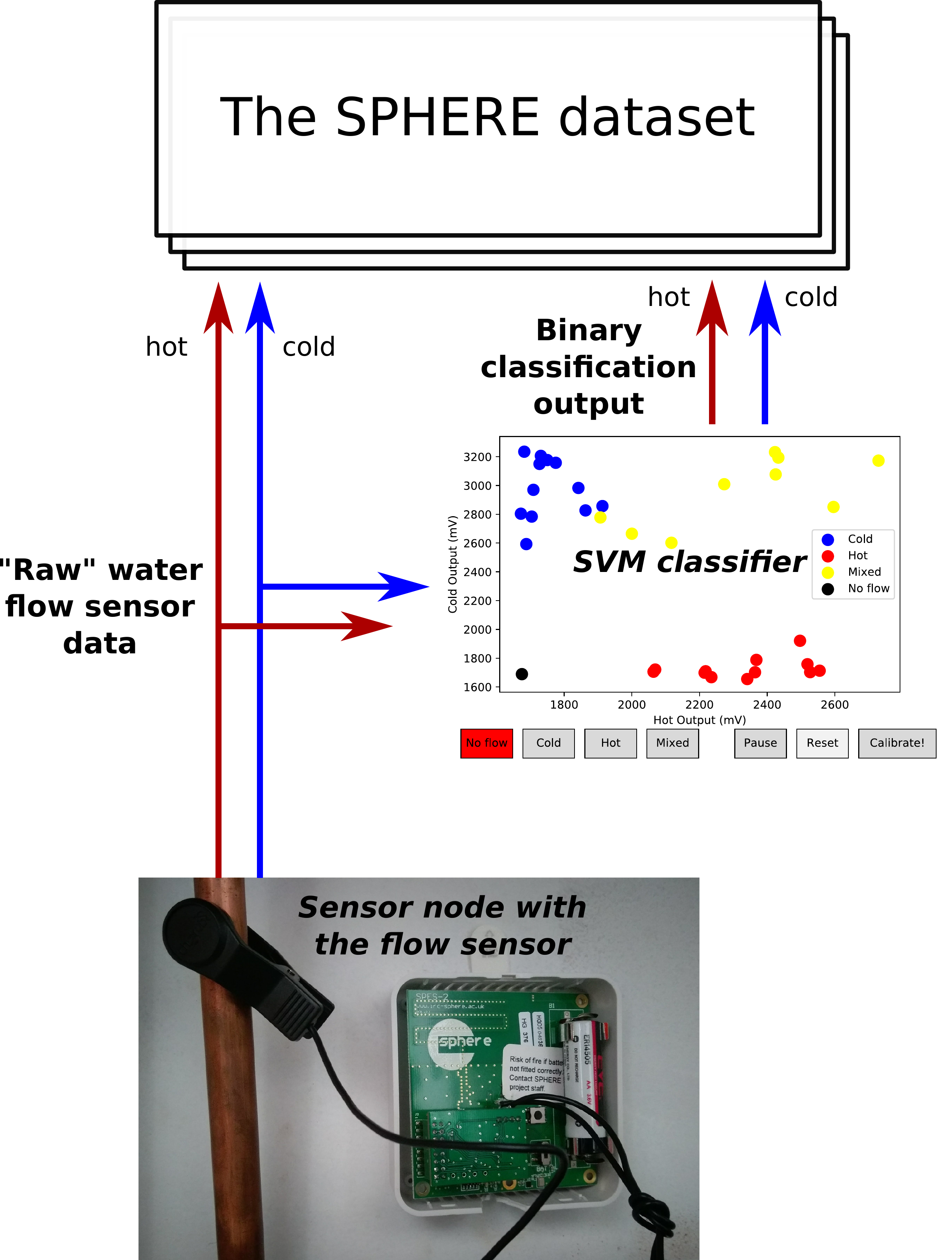}
  \caption{Overview of the water flow sensing subsystem~\cite{vafeas2018}. A piezoelectric sensor is clamped on the water pipe, near the water tap; there is one clamp for the hot water pipe, and one for the cold one. The sensor data is delivered to the SPHERE Home Gateway, where the water flow status is classified with a machine learning algorithm. Both the raw flow values, the classification results, and the calibration data of the classifier are saved in the dataset.}
  \label{fig:watersensor}
\end{figure}

Raw flow data example:

\begin{lstlisting}[language=json]
  {
	'_id': ObjectId('5a005075b190070a60eec8cf'),
	'e': [{
		'n': 'PIEZO_WATER_C',
		'v': 13372608
	}, {
		'n': 'PIEZO_WATER_H',
		'v': 13380796
	}, {
		'n': 'LT',
		'v': 0
	}],
	'bt': datetime.datetime(2017, 11, 6, 12, 6, 59, 263000),
	'ts': 80552007,
	'uid': 'fd00::212:4b00:0:ff80',
	'tso': 1509164499.1936808,
	'mc': 94846,
	'hid': '0000'
  }
  \end{lstlisting}

The architecture of the water flow sensing is given in Fig.~\ref{fig:watersensor}; for more details consult~\cite{vafeas2018}.

  \begin{itemize}
  \item     \code{PIEZO\_WATER\_C} -- a 32-bit integer value encapsulating start of reading value (12 bits), end of reading value(12 bits) and the interrupt status (1 bit) of the \textbf{cold} water flow sensor.
  \item     \code{PIEZO\_WATER\_H} -- a 32-bit integer value encapsulating start of reading value, end of reading value and the interrupt status of the \textbf{hot} water flow sensor. The values are encapsulated in the same way as the values for the cold flow.
  \end{itemize}

\linesep
Flow classification example:

\begin{lstlisting}[language=json]
  {
	'_id': ObjectId('5a005075b190070a60eec8cd'),
	'uid': 'fd00::212:4b00:0:ff80',
	'bt': datetime.datetime(2017, 11, 6, 12, 6, 59, 263000),
	'tso': 1509164499.1936808,
	'e': [{
		'n': 'WATER_COLD_ON',
		'v': False
	}, {
		'n': 'WATER_HOT_ON',
		'v': False
	}],
	'ts': 80552007,
	'hid': '4954'
  }
\end{lstlisting}
  
  \begin{itemize}
  \item \code{WATER\_COLD\_ON} -- a Boolean value describing whether the \textbf{cold} water flow is classified as ``on''.
  \item \code{WATER\_HOT\_ON} -- a Boolean value describing whether the \textbf{hot} water flow is classified as ``on''.
  \end{itemize}
These values are only produced if the water sensor calibration process has been successfully completed (each SPHERE house requires separate calibration process).

\subsubsection{Appliance monitor data}

SPHERE uses the common-off-the-shelf CurrentCost Individual Appliance Monitors (IAMs)\footnote{http://www.currentcost.com/product-iams-specification.html} and a mains meter sensor, called the Energy Transmitter \footnote{http://www.currentcost.com/product-transmitter-specifications.html}, which every 6-7 seconds communicate their energy usage data to the NetSmart gateway \footnote{http://www.currentcost.com/product-netsmart-specification.html}. The CurrentCost gateway is connected via a USB-serial connection using an RJ45 connector on the gateway end. Data transmitted through the serial port to the SHG is picked up by a script that parses and formats each XML packet into JSON and forwards the traffic onto the MQTT queue. Each data packet has some additional information inserted, such as \textit{hid}, and the \textit{bt} of the SHG.
\\
Electricity usage message example:
\begin{lstlisting}[language=json]
{ 
    "uid" : "02052", 
    "bt" : ISODate("2018-04-24T13:28:19.014+0000"), 
    "e" : [
        {
            "n" : "ELEC", 
            "v" : "00013"
        }
    ], 
    "hid" : "9999"
}
\end{lstlisting}

\begin{itemize}
  \item Per device:
    \begin{itemize}
    \item \code{n} -- sensor name i.e. 'ELEC'.
    \item \code{v} -- energy usage value in Watts. This string is actually of integer type of length 5 with leading zeros. The example above denotes an energy consumption of 13 Watts. 
    \end{itemize}
\end{itemize}
    
Packet timestamps are assigned by the SHG since CurrentCost does not offer an accurate clock solution for their network -- time is reported to a seconds level accuracy and needs to be set manually on the gateway.

\subsection{Wearable data}

Wearable message example:
\begin{lstlisting}[language=json]
{
	'_id': ObjectId('5a00512ab190070a60eed598'),
	'e': [{
		'n': 'ACCEL',
		't': 0.4,
		'v': [0.096, -0.864, -0.064]
	}, {
		'n': 'ACCEL',
		't': 0.32,
		'v': [-0.096, -0.96, -0.192]
	}, {
		'n': 'ACCEL',
		't': 0.24,
		'v': [-0.064, -0.96, -0.064]
	}, {
		'n': 'ACCEL',
		't': 0.16,
		'v': [-0.256, -0.832, -0.256]
	}, {
		'n': 'ACCEL',
		't': 0.08,
		'v': [-0.256, -1.024, -0.224]
	}, {
		'n': 'ACCEL',
		't': 0,
		'v': [-0.064, -0.832, -0.032]
	}],
	'bt': datetime.datetime(2017, 11, 6, 12, 9, 59),
	'gw': [{
		'uid': 'fd00::212:4b00:0:ff05',
		'rssi': -89,
		'mc': 171761,
		'ts': 80569981
	}, {
		'uid': 'fd00::212:4b00:0:ff03',
		'rssi': -84,
		'mc': 1691039,
		'ts': 80569982
	}, {
		'uid': 'fd00::212:4b00:0:ff04',
		'rssi': -82,
		'mc': 2119128,
		'ts': 80569982
	}, {
		'uid': 'fd00::212:4b00:0:ff06',
		'rssi': -84,
		'mc': 2621139,
		'ts': 80569982
	}, {
		'uid': 'fd00::212:4b00:0:ff07',
		'rssi': -76,
		'mc': 1885067,
		'ts': 80569983
	}],
	'ts': 80569981,
	'uid': 'a0:e6:f8:00:ff:c0',
	'tso': 1509164499.1903248,
	'mc': 44539,
	'hid': '0000'
}
\end{lstlisting}

  \begin{itemize}
  \item \code{ACCEL} -- acceleration data
    \begin{itemize}
    \item \code{v} -- the x, y, z coordinates of the acceleration data (always in this order)
    \item \code{t} -- the timestamp offset (in seconds) of this data relative to the timestamp of the document.
    \end{itemize}
  \end{itemize}
  Packet timestamps are assigned by gateways, not wearables, since wearables are not time-synchronized with the rest of the SPHERE's network.
  
  In a wearable data document, the top-level \code{ts} field is equal to the smallest of the \code{ts} fields inside the \code{gw} array: that is, we treat the timestamp assigned by the gateway that ``first'' received the packet as the timestamp of the packet. In the example, the timestamps of the different gateways are slightly different: the maximal timestamp is $0.02$ seconds larger than the minimal timestamp. This is a very common occurrence; an error of this magnitude is present in $\approx 30$ percent on wearable packets (see the ``data quality'' section for detailed discussion).

  In contrast to timestamps, there are multiple, independent sequence numbers in the wearable data document. The top-level \code{mc} field refers to the sequence number of the wearable that generated the acceleration data.
  In theory, this field can be used to restore higher accuracy timestamps of the wearable data in case the \code{bt} field is deemed insufficiently accurate.
  The \code{mc} fields in the elements of the \code{gw} array refer to the sequence numbers of the gateways that forwarded the data.

\subsection{Video data}
The video data is constituted by two different types of messages: the bounding box information and the actual silhouette image. Each silhouette image entry is characterized by the first element of \textit{e} being the text ``silhouette''. This entry contains the png file of the raw silhouette encoded in form of base64. All the entries following the silhouette are bounding boxes, detected by the person detector. To collect all the bounding boxes detected in a specific frame, it is useful to query the database for elements which have the same base time \textit{bt} and the same \textit{uid} (the camera room id) as the specified silhouette frame.

Video message example containing a silhouette$^*$:
\begin{lstlisting}[language=json]
{
  '_id': ObjectId('59d4f16eb159dd06d20053ac'),
  'bt': datetime.datetime(2017, 10, 4, 14, 34, 22, 683000),
  'e': [{
    'n': 'silhouette', 
    'v': 'iVBORw0KGgoAAAANSUhEUgAAAoAAAAHgCAAAAAAQuoM4AAAPFElEQVR4Ae3Bi3aiQABEwdtz+P9f7tUYN5r4AISZAbqKiIiIiIiIi ...' # base64 format
  }], 
  'hid': u'4954',
  'uid': u'b8aeed7edc31'
}
\end{lstlisting}

$^*$ To represent the silhouettes, the Base64 format is used. Base64 is a group of similar binary-to-text encoding schemes that represent binary data in an ASCII string format by translating it into a radix-64 representation. The term Base64 originates from a specific MIME content transfer encoding. Each base64 digit represents exactly 6 bits of data. Three 8-bit bytes (i.e., a total of 24 bits) can therefore be represented by four 6-bit base64 digits.

Video message example for a bounding box:
\begin{lstlisting}[language=json]
{
  '_id': ObjectId('59d9db1ab190070a60c977c3'),
  'bt': datetime.datetime(2017, 10, 8, 8, 0, 26, 912000),
  'e': [{
    'n': 'frameID', 
    'v': 123178
  }, {
    'n': 'userID', 
    'v': 1089350673
  }, {
    'n': '2Dbb', 
    'v': [500, 96, 595, 263]
  }, {
    'n': '2DCen', 
    'v': [554, 151]
  },  {
    'n': '3Dbb',
    'v': [1430, 1144, 4533, 2475, -207, 5133]
  }, {
    'n': '3Dcen', 
    'v': [1989, 748, 4833]
  }, {
    'n': 'Activity', 
    'v': u'Sitting Down'
  }, {
    'n': 'Intensity', 
    'v': u'Light+'
  }, {
    'n': 'FeaturesREID', 
    'v': [0.4997463822364807, -0.7747822403907776, -2.420274496078491, 0.8493637442588806, 0.151933953166008, 1.3922899961471558, ...]
  }],
  'hid': '4954',
  'uid': 'b8aeed7edc31'}
\end{lstlisting}

\begin{itemize}
	\item \code{frameID} -- Frame ID
	\begin{itemize}
		\item \code{v} -- frame ID number.
	\end{itemize}
    \item \code{userID} -- User ID
	\begin{itemize}
		\item \code{v} -- user ID number. Used for tracking bounding boxes while the person appear in the frame
	\end{itemize}
    \item \code{2Dbb} -- 2D bounding box
	\begin{itemize}
		\item \code{v} -- Horizontal and vertical coordinates of the bottom-left and top-right corners
	\end{itemize}
    \item \code{2DCen} -- Center of the 2D bounding box
	\begin{itemize}
		\item \code{v} -- 2D bounding box center pixel coordinates. Calculated from \code{2Dbb}
	\end{itemize}
    \item \code{3Dbb} -- 3D bounding box
	\begin{itemize}
		\item \code{v} -- Horizontal, vertical and depth coordinates of the bottom-left-front and top-right-back corners 3D bounding box
	\end{itemize}
    \item \code{3Dcen} -- Center of the 3D bounding box. Calculated from \code{3Dbb}
	\begin{itemize}
		\item \code{v} -- 3D bounding box center pixel coordinates
	\end{itemize}
    \item \code{Activity} -- Activity label (not yet validated)
	\begin{itemize}
		\item \code{v} -- string that describe the type of activity
	\end{itemize}
    \item \code{Intensity} -- Intensity value (not yet validated)
	\begin{itemize}
		\item \code{v} -- value of the activities' intensity 
	\end{itemize}
    \item \code{FeaturesREID} -- Features for Re-Identification 
	\begin{itemize}
		\item \code{v} -- Feature vector (dimension 546) to perform person re-identification and support other type of classification tasks, which consists of:
        \begin{itemize}
          \item 500 projection coefficients which is combination of overlapping colour and texture features.
          \item 45 illumination invariant colour space features.
          \item 1 value that is an estimate of head size.
        \end{itemize}
        This entry only exists when the subject is fully captured in the frame and it is at the right distance from the camera.
	\end{itemize}
\end{itemize}

\subsection{Monitoring data}

\textbf{Example of a heartbeat message from the gateway script (with main gateway node connected).} This is from a test network where gateway nodes \texttt{01} and \texttt{03} and environmental sensor node \texttt{80} connected and sending messages. \texttt{03} is missing \texttt{ACCEL}, i.e. wearable data. Wearable \texttt{c0} is also present, but its data is only received by the Video Gateway NUC \texttt{b8:ae:ed:e9:d3:c0} and Home Gateway NUC \texttt{f4:4d:30:6c:66:93}, not by the SPG-2 gateway.

\begin{lstlisting}[language=json]
{
	'_id': ObjectId('5aba16e07ea12b8f29f0be22'),
	'e': [
		[{
			'n': 'UID',
			'v': 'fd00::212:4b00:0:fa03'
		}, {
			'n': 'MODEL',
			'v': 'SPG2_F'
		}, {
			'n': 'NUM_MESSAGES',
			'v': 1
		}, {
			'n': 'NUM_TOTAL_MESSAGES',
			'v': 488659
		}, {
			'n': 'LOCATION',
			'v': 'sphere office 1'
		}, {
			'n': 'RSSI',
			'v': -30
		}, {
			'n': 'LAST_MESSAGE',
			'v': 1522144946.720782
		}, {
			'n': 'MISSING_SENSORS',
			'v': ['ACCEL']
		}],
		[{
			'n': 'UID',
			'v': 'fd00::212:4b00:0:fa80'
		}, {
			'n': 'MODEL',
			'v': 'SPES2'
		}, {
			'n': 'NUM_MESSAGES',
			'v': 2
		}, {
			'n': 'NUM_TOTAL_MESSAGES',
			'v': 39768
		}, {
			'n': 'LOCATION',
			'v': 'sphere office 1'
		}, {
			'n': 'RSSI',
			'v': -20
		}, {
			'n': 'LAST_MESSAGE',
			'v': 1522144964.9395883
		}],
		[{
			'n': 'UID',
			'v': 'fd00::212:4b00:0:fa01'
		}, {
			'n': 'MODEL',
			'v': 'SPG2_BR'
		}, {
			'n': 'NUM_MESSAGES',
			'v': 1
		}, {
			'n': 'NUM_TOTAL_MESSAGES',
			'v': 34011
		}, {
			'n': 'LOCATION',
			'v': 'sphere office  1'
		}, {
			'n': 'LAST_MESSAGE',
			'v': 1522144961.2643692
		}],
		[{
			'n': 'UID',
			'v': 'a0:e6:f8:00:fa:c0'
		}, {
			'n': 'MODEL',
			'v': 'SPW2'
		}, {
			'n': 'NUM_MESSAGES',
			'v': 494
		}, {
			'n': 'NUM_TOTAL_MESSAGES',
			'v': 7961121
		}, {
			'n': 'LOCATION',
			'v': 'A'
		}, {
			'n': 'LAST_MESSAGE',
			'v': 1522144987.4976475
		}],
		[{
			'n': 'UID',
			'v': 'f4:4d:30:6c:66:93'
		}, {
			'n': 'MODEL',
			'v': 'Intel NUC HOME'
		}, {
			'n': 'NUM_MESSAGES',
			'v': 248
		}, {
			'n': 'NUM_TOTAL_MESSAGES',
			'v': 3542351
		}, {
			'n': 'LOCATION',
			'v': 'sphere office 1'
		}, {
			'n': 'LAST_MESSAGE',
			'v': 1522144986.548605
		}],
		[{
			'n': 'UID',
			'v': 'b8:ae:ed:e9:d3:c0'
		}, {
			'n': 'MODEL',
			'v': 'Intel NUC VIDEO'
		}, {
			'n': 'NUM_MESSAGES',
			'v': 246
		}, {
			'n': 'NUM_TOTAL_MESSAGES',
			'v': 3493470
		}, {
			'n': 'LOCATION',
			'v': 'sphere office - hall 1'
		}, {
			'n': 'LAST_MESSAGE',
			'v': 1522144987.4976475
		}], {
			'n': 'UPTIME',
			'v': 926114.7416009903
		}, {
			'n': 'SERIAL_PORT',
			'v': 1
		}, {
			'n': 'GW_NODE_ALIVE',
			'v': 1
		}
	],
	'uid': 'gateway_script',
	'bt': '2018-03-27T10:03:08.380Z',
	'hid': '0250',
	'Date': '2018-03-27T10:03:08.384Z',
	'type': 'HEARTBEAT'
}
\end{lstlisting}

\begin{itemize}
  \item Per device:
    \begin{itemize}
    \item \code{UID} -- ID of the device.
    \item \code{LOCATION} -- information about the device's location, if present in HyperCat.
    \item \code{MODEL} -- the model of the device.
    \item \code{NUM\_MESSAGES} -- number of messages from the device in the last reporting period (1~min).
    \item \code{NUM\_TOTAL\_MESSAGES} --number of messages from the device since the start of the script.
    \item \code{LAST\_MESSAGE} -- the local time when last message was received from the device.
    \item \code{ERRORS} -- any specific errors for that node (e.g. invalid environmental sensor readings) since the last heartbeat message.
    \item \code{NEXTHOP} -- the node through which this device is connected (only for TSCH devices).
    \item \code{RSSI} -- the RSSI of the last packet from the device (only for directly connected TSCH devices).
    \item \code{MISSING\_SENSORS} -- array of sensors detected as not reported by the device, if any. For example, having \code{ACCEL} here is normal in case the wearables are not in the communication range of the device.
    \end{itemize}
  \item \code{SERIAL\_PORT} -- whether the serial port is open.
  \item \code{GW\_NODE\_ALIVE} -- whether the connected node is communicating.
  \item \code{UPTIME} -- the uptime of the script.
  \end{itemize}

\linesep
\textbf{Example of a heartbeat message from the gateway script (with backup gateway node connected):}

\begin{lstlisting}[language=json]
{
	'_id': ObjectId('5aba17027ea12b8f29f0bed9'),
	'e': [{
		'n': 'UPTIME',
		'v': 926147.5098118782
	}, {
		'n': 'SERIAL_PORT',
		'v': 1
	}, {
		'n': 'GW_NODE_ALIVE',
		'v': 1
	}],
	'uid': 'gateway_script',
	'bt': '2018-03-27T10:03:41.148Z',
	'hid': '0250',
	'Date': '2018-03-27T10:03:41.153Z',
	'type': 'HEARTBEAT'
}
\end{lstlisting}

\linesep
Packet delivery rate message, gateway script (with main gateway node connected):

\begin{lstlisting}[language=json]
{
	'_id': ObjectId('5abc1bbf7ea12b8f29fe08b9'),
	'e': [
		[{
			'v': 'fd00::212:4b00:0:fa01',
			'n': 'UID'
		}, {
			'v': 0,
			'n': 'PDR_PRESENT'
		}, {
			'v': 0,
			'n': 'PDR_TOTAL'
		}],
		[{
			'v': 'fd00::212:4b00:0:fa03',
			'n': 'UID'
		}, {
			'v': 83,
			'n': 'PDR_PRESENT'
		}, {
			'v': 83,
			'n': 'PDR_TOTAL'
		}],
		[{
			'v': 'fd00::212:4b00:0:fa80',
			'n': 'UID'
		}, {
			'v': 158,
			'n': 'PDR_PRESENT'
		}, {
			'v': 158,
			'n': 'PDR_TOTAL'
		}],
		[{
			'v': 'a0:e6:f8:00:fa:c0',
			'n': 'UID'
		}, {
			'v': 13858,
			'n': 'PDR_PRESENT'
		}, {
			'v': 14429,
			'n': 'PDR_TOTAL'
		}],
		[{
			'v': 'serial',
			'n': 'UID'
		}, {
			'v': 1269,
			'n': 'NUM_MESSAGES'
		}, {
			'v': 1942,
			'n': 'NUM_TOTAL_MESSAGES'
		}]
	],
	'uid': 'gateway_script',
	'mc': 35,
	'bt': '2018-03-28T22:48:25.910Z',
	'hid': '0250',
	'Date': '2018-03-28T22:48:25.913Z',
	'type': 'PDR'
}
\end{lstlisting}

  \begin{itemize}
  \item \code{PDR\_PRESENT} -- how many unique messages were received from the device.
  \item \code{PDR\_PRESENT} -- how many unique messages did the device sent (an estimate using the ``message count'' field).
  \end{itemize}
These statistics refer to 1~hour large time period. Each PDR statistics message with same data is published multiple times to MQTT increase its chances of getting through bad 3G connections. In the database, duplicate messages can be recognized using the \code{mc} field.

\linesep
Packet delivery rate message, gateway script (with backup gateway node connected):

\begin{lstlisting}[language=json]
{
	'_id': ObjectId('5abbb9447ea12b8f29fbaf89'),
	'e': [
		[{
			'v': 'serial',
			'n': 'UID'
		}, {
			'v': 0,
			'n': 'NUM_MESSAGES'
		}, {
			'v': 0,
			'n': 'NUM_TOTAL_MESSAGES'
		}]
	],
	'uid': 'gateway_script',
	'mc': 28,
	'bt': '2018-03-28T15:48:13.985Z',
	'hid': '0250',
	'Date': '2018-03-28T15:48:13.993Z',
	'type': 'PDR'
}
\end{lstlisting}

\linesep
Heartbeat message, BLE script:

\begin{lstlisting}[language=json]
{
	'_id': ObjectId('5ab989d57ea12b8f29eca7a1'),
	'e': [{
		'n': 'NUM_MESSAGES_SINCE_LAST',
		'v': 1576
	}, {
		'n': 'NUM_MESSAGES_TOTAL',
		'v': 4562
	}],
	'uid': 'BLE_SCRIPT_b8aeed7db949',
	'bt': '2018-03-27T00:01:20.999Z',
	'hid': '0000',
	'Date': '2018-03-27T00:01:20.998Z',
	'type': 'HEARTBEAT'
}
\end{lstlisting}
  \begin{itemize}
  \item \code{NUM\_MESSAGES\_TOTAL} -- number of wearable messages received since the start of the script.
  \item \code{NUM\_MESSAGES\_SINCE\_LAST} -- number of wearable messages received since the last report.
  \end{itemize}

\linesep
Monitoring message, SPG-2 and SPES-2 nodes:

\begin{lstlisting}[language=json]
{
 	'_id': ObjectId('5a039ae7b190070a602a721e'),
 	'uid': 'fd00::212:4b00:0:ff87',
 	'bt': datetime.datetime(2017, 11, 9, 0, 1, 43, 924000),
 	'tso': 0,
 	'e': [{
 		'n': 'UPTIME',
 		'v': 2908940
 	}, {
 		'n': 'FLASH_STATUS',
 		'v': 1
 	}, {
 		'n': 'PROCESS_MAX_EVENTS',
 		'v': 3
 	}, {
 		'n': 'REBOOT_REASON',
 		'v': 1
 	}, {
 		'n': 'REBOOT_IS_DELIBERATE',
 		'v': 1
 	}, {
 		'n': 'NUM_REBOOTS',
 		'v': 17
 	}, {
 		'n': 'MAX_STACK_SIZE',
 		'v': 1530
 	}],
 	'ts': 151018570392.4064,
 	'hid': '0000',
 	'Date': datetime.datetime(2017, 11, 9, 0, 1, 43, 925000)
}
\end{lstlisting}
 \begin{itemize}
  \item \code{UPTIME} -- uptime of the device.
  \item \code{FLASH\_STATUS} -- nonzero if whether external flash read/write test was successful, zero otherwise.
  \item \code{PROCESS\_MAX\_EVENT} -- maximal number of events in the Contiki process event buffer.
  \item \code{MAX\_STACK\_SIZE} -- the maximal size of stack used by the application during the execution so far.
  \item \code{REBOOT\_REASON} -- the reason code for the last reboot. Consult Texas Instruments CC2650 documentation for the meaning of these codes.
  \item \code{REBOOT\_IS\_DELIBERATE} -- nonzero if the last reboot was initialized by the software of the node, zero otherwise. Reboots are initialized when problems are detected, e.g. no CoAP requests are received for 2 hours.
  \item \code{NUM\_REBOOTS} -- the total number of reboots in the devices life cycle so far.
  \end{itemize}

\linesep
Battery voltage monitoring message, SPES-2 nodes:
\begin{lstlisting}[language=json]
{
	'_id': ObjectId('5a03c5d4b190070a602c9207'),
	'uid': 'fd00::212:4b00:0:ff85',
	'bt': datetime.datetime(2017, 11, 9, 3, 4, 52, 52000),
	'tso': 1509164487.6929398,
	'e': [{
		'n': 'BATMON_VOLT',
		'v': 3.585
	}],
	'ts': 103220436,
	'hid': '0000',
	'Date': datetime.datetime(2017, 11, 9, 3, 4, 52, 361000)
}
\end{lstlisting}
\begin{itemize}
  \item \code{BATMON\_VOLT} -- battery voltage of the device.
\end{itemize}

\linesep
Monitoring message, wearable node:

\begin{lstlisting}[language=json]
{
	'_id': ObjectId('5ab989d57ea12b8f29eca79f'),
	'tso': 1520032700.7959356,
	'e': [{
		'n': 'BATMON_VOLT',
		'v': 3.333
	}, {
		'n': 'UPTIME',
		'v': 4937677
	}],
	'uid': 'a0:e6:f8:00:ff:c2',
	'ts': 207618107,
	'bt': '2018-03-27T00:01:21.865Z',
	'hid': '0000',
	'Date': '2018-03-27T00:01:22.325Z'
}
\end{lstlisting}
\begin{itemize}
  \item \code{UPTIME} -- uptime of the device in seconds.
  \item \code{BATMON\_VOLT} -- battery voltage of the device in volts.
  \item \code{AXLCONF} -- accelerometer configuration (\textit{i.e.} the filter control register, \texttt{FILTER\_CTL} \cite{ADXL362}).
  \item \code{AXLSTAT} -- accelerometer status (\textit{i.e.} the status register, \texttt{STATUS} \cite{ADXL362}).
  \item \code{ERR} -- error code, shown below.
  \item \code{CPUTEMP} -- CPU temperature in $^o$C.
  \item \code{CPUVDD} -- CPU voltage in volts.
  \item \code{SW\_VERSION} -- software revision number.
  \item \code{HW\_VERSION} -- hardware revision number (board revision number multiplied by $10$).
  \end{itemize}
  
\linesep
Monitoring message, wearable error codes:
\begin{lstlisting}[language=c]
typedef enum
{
	ERROR_NO_ERROR=0,       // Everything OK
	ERROR_AXL0_DEAD,        // Accelerometer 0 not responding
	ERROR_AXL0_FIFO,        // Accelerometer 0 unexpected number of samples in FIFO
	ERROR_AXL0_FIFO_FULL,   // Accelerometer 0 has full FIFO
	ERROR_AXL1_DEAD,        // Accelerometer 1 not responding
	ERROR_AXL1_FIFO,        // Accelerometer 1 unexpected number of samples in FIFO
	ERROR_SPI0_FAIL,        // Failed to initialize SPI0
	ERROR_SPI1_FAIL,        // Failed to initialize SPI1
	ERROR_FLASH,            // Failed to initialize external flash memory
	ERROR_ADV_CONFIG,       // Failed to configure accelerometer
	ERROR_CRYPTO_INIT,      // Failed to initialize crypto engine
	ERROR_CRYPTO_OP         // Failed to encrypt or decrypt
} ERROR_TYPE; // Wearable Error Type
\end{lstlisting}

\linesep
TSCH network's RSSI strength message:

\begin{lstlisting}[language=json]
{
	'_id': ObjectId('5a60923fb190070a60fdd63c'),
	'uid': 'fd00::212:4b00:0:ff04',
	'bt': datetime.datetime(2018, 1, 18, 12, 25, 35, 173000),
	'tso': 0,
	'e': [{
		'n': 'RSSI',
		'v': -71
	}, {
		'n': 'CHANNEL',
		'v': 25
	}],
	'ts': 151627833517.32825,
	'hid': '0000',
	'Date': datetime.datetime(2018, 1, 18, 12, 25, 35, 174000)
}
\end{lstlisting}
  \begin{itemize}
  \item \code{RSSI} -- signal strength value.
  \item \code{CHANNEL} -- IEEE~802.15.4 channel on which the message is received.
  \end{itemize}

\linesep
Link-layer neighbor status message (produced for each active neighbor in the neighbor table):

\begin{lstlisting}[language=json]
{
	'_id': ObjectId('5a608e9cb190070a60fda9ce'),
	'uid': 'fd00::212:4b00:0:ff80',
	'bt': datetime.datetime(2018, 1, 18, 12, 10, 4, 468000),
	'tso': 0,
	'e': [{
		'n': 'LINK_NEIGHBOR',
		'v': 'fd00::212:4b00:0:ff05'
	}, {
		'n': 'LINK_ETX',
		'v': 128
	}, {
		'n': 'LINK_PACKETS_TX_PREVIOUS_HOUR',
		'v': 671
	}, {
		'n': 'LINK_PACKETS_ACK_PREVIOUS_HOUR',
		'v': 541
	}, {
		'n': 'LINK_PACKETS_RX_PREVIOUS_HOUR',
		'v': 2337
	}],
	'ts': 151627740446.8441,
	'hid': '0000',
	'Date': datetime.datetime(2018, 1, 18, 12, 10, 4, 472000)
}
\end{lstlisting}
\begin{itemize}
\item \code{LINK\_NEIGHBOR} -- the IPv6 address of the neighbor.
  \item \code{LINK\_ETX} -- EWMA of the Expected Transmission Count to the neighbor, multiplied by 128.
  \item \code{LINK\_PACKETS\_TX\_PREVIOUS\_HOUR} -- number of packets transmitted (including duplicates).
  \item \code{LINK\_PACKETS\_ACK\_PREVIOUS\_HOUR} -- number of packets ACKed.
  \item \code{LINK\_PACKETS\_RX\_PREVIOUS\_HOUR} -- number of packets received.
  \end{itemize}
  See Contiki~3.x documentation and source code for more details on the ETX value.

\linesep
TSCH node's status message:

\begin{lstlisting}[language=json]
{
	'_id': ObjectId('5a608ea0b190070a60fdaa0f'),
	'uid': 'fd00::212:4b00:0:ff80',
	'bt': datetime.datetime(2018, 1, 18, 12, 10, 8, 318000),
	'tso': 1513623272.9585357,
	'e': [{
		'n': 'TSCH_OUTGOING_DROPPED',
		'v': 0
	}, {
		'n': 'TSCH_INCOMING_DROPPED',
		'v': 0
	}, {
		'n': 'TSCH_MAX_SYNC_ERROR',
		'v': 229
	}, {
		'n': 'TSCH_HOPPING_SEQUENCE',
		'v': [12, 14, 13, 25, 20, 11, 16]
	}, {
		'n': 'TSCH_CHANNEL_IDLE_RSSI',
		'v': [0, 0, 0, 0, 0, 0, 0, 0, 0, 0, 0, 0, 0, 0, 0, 0]
	}, {
		'n': 'TSCH_TIMESOURCE',
		'v': 'fd00::212:4b00:0:ff01'
	}, {
		'n': 'TSCH_DRIFT',
		'v': 8.953
	}],
	'ts': 265413536,
	'hid': '0000',
	'Date': datetime.datetime(2018, 1, 18, 12, 10, 8, 429000)
}
\end{lstlisting}
  \begin{itemize}
  \item \code{TSCH\_OUT\_OK} -- number of outgoing packets transmitted successfully.
  \item \code{TSCH\_OUTGOING\_DROPPED} -- number of outgoing packets dropped due to all reasons. Present only in SPHERE version D; version E breaks down this to a more detailed information (see further).
  \item \code{TSCH\_OUT\_DROPPED\_TX\_LIMIT} -- number of outgoing packets dropped due to exceeded retransmission count.
  \item \code{TSCH\_OUT\_DROPPED\_QUEUE\_FULL} -- number of outgoing packets dropped due to full queue.
  \item \code{TSCH\_OUT\_DROPPED\_NO\_ROUTE} -- number of outgoing packets dropped due to not having route to the destination.
  \item \code{TSCH\_OUT\_DROPPED\_NO\_NEIGHBOR} -- number of outgoing packets dropped due to not having entry for the nexthop in the neighbor tables.
  \item \code{TSCH\_OUT\_DROPPED\_OTHER} -- number of outgoing packets dropped due to other reasons.
  \item \code{TSCH\_INCOMING\_DROPPED} -- number of incoming packets dropped.
  \item \code{TSCH\_NUM\_DISSOCIATIONS} -- number of times the node has left the TSCH network since the last reboot.
  \item \code{TSCH\_MAX\_SYNC\_ERROR} -- maximal time synchronization error, in microseconds.
   
  \item \code{TSCH\_HOPPING\_SEQUENCE} -- the current hopping sequence.
  \item \code{TSCH\_CHANNEL\_IDLE\_RSSI} -- background noise RSSI.
  \item \code{TSCH\_TIMESOURCE} -- TSCH time source node's address (same as RPL routing parent).
  \item \code{TSCH\_DRIFT} -- clock drift estimate, in ppm.
  \end{itemize}
\code{TSCH\_CHANNEL\_IDLE\_RSSI} has valid readings only on the root gateway node.

\linesep
TSCH node's neighbor status message (produced only for the timesource neighbor):

\begin{lstlisting}[language=json]
{
	'_id': ObjectId('5a608ea4b190070a60fdaa29'),
	'uid': 'fd00::212:4b00:0:ff80',
	'bt': datetime.datetime(2018, 1, 18, 12, 10, 12, 383000),
	'tso': 0,
	'e': [{
		'n': 'TSCH_NEIGHBOR',
		'v': 'fd00::212:4b00:0:3301'
	}, {
		'n': 'TSCH_HOPPING_SEQUENCE',
		'v': [12, 14, 13, 25, 20, 11, 16]
	}, {
		'n': 'TSCH_RSSI',
		'v': [-73, -74, -74, -77, -76, -75, -77]
	}, {
		'n': 'TSCH_LQI',
		'v': [57, 58, 58, 56, 58, 57, 58]
	}, {
		'n': 'TSCH_P_TX',
		'v': [0.92, 1, 1, 1, 1, 1, 0.99]
	}],
	'ts': 151627741238.31018,
	'hid': '0000',
	'Date': datetime.datetime(2018, 1, 18, 12, 10, 12, 384000)
}
\end{lstlisting}
  \begin{itemize}
  \item \code{TSCH\_NEIGHBOR} -- the neighbor's IPv6 address.
  \item \code{TSCH\_HOPPING\_SEQUENCE} -- the TSCH hopping sequence currently used.
  \item \code{TSCH\_RSSI} -- per-channel RSSI values for packets received from that neighbor. EWMA-filtered.
  \item \code{TSCH\_LQI} -- per-channel LQI values for packets received from that neighbor. EWMA-filtered.
  \item \code{TSCH\_P\_TX} -- the $P_{tx}$, successful packet acknowledgement probability from that neighbor. EWMA-filtered values for the channels in the TSCH hopping sequence.
  \end{itemize}
  Note that these statistics are accurate only for channels that have been used for a sufficiently long time.

\linesep
Contiki energest status message:

\begin{lstlisting}[language=json]
{
	'_id': ObjectId('5a608eadb190070a60fdaab2'),
	'uid': 'fd00::212:4b00:0:ff80',
	'bt': datetime.datetime(2018, 1, 18, 12, 10, 21, 298000),
	'tso': 0,
	'e': [{
		'n': 'ENERGEST_TYPE_OPT',
		'v': 499756858
	}, {
		'n': 'ENERGEST_TYPE_FLASH_STANDBY',
		'v': 1482
	}, {
		'n': 'ENERGEST_TYPE_LED_RED',
		'v': 2794
	}, {
		'n': 'ENERGEST_TYPE_FLASH_ERASE',
		'v': 0
	}, {
		'n': 'ENERGEST_TYPE_FLASH_WRITE',
		'v': 0
	}, {
		'n': 'ENERGEST_TYPE_HDC',
		'v': 94983338
	}, {
		'n': 'ENERGEST_TYPE_LISTEN',
		'v': 6477566
	}, {
		'n': 'ENERGEST_TYPE_LED_YELLOW',
		'v': 5698
	}, {
		'n': 'ENERGEST_TYPE_BMP',
		'v': 95860702
	}, {
		'n': 'ENERGEST_TYPE_FLASH_READ',
		'v': 1318
	}, {
		'n': 'ENERGEST_TYPE_TRANSMIT',
		'v': 0
	}, {
		'n': 'ENERGEST_TYPE_LED_GREEN',
		'v': 5710
	}, {
		'n': 'ENERGEST_TYPE_DEEP_LPM',
		'v': 2711035296.0
	}, {
		'n': 'ENERGEST_TYPE_LPM',
		'v': 402464618
	}, {
		'n': 'ENERGEST_TYPE_CPU',
		'v': 122434880
	}],
	'ts': 151627742129.8406,
	'hid': '0000',
	'Date': datetime.datetime(2018, 1, 18, 12, 10, 21, 300000)
}
\end{lstlisting}
Note: this describes \textbf{only} the energy consumption of the system outside TSCH timeslot processing! Radio Tx/Rx time should be inferred for packet Tx/Rx statistics instead.

  \begin{itemize}
  \item \code{ENERGEST\_TYPE\_DEEP\_LPM} -- number of \code{rtimer} ticks spent in deep-sleep mode ($0.016\,\mu$A average current consumption on SPES-2 devices).
  \item \code{ENERGEST\_TYPE\_LPM} -- number of \code{rtimer} ticks spent in sleep mode ($1.335\,\mu$A average current consumption on SPES-2 devices).
  \item \code{ENERGEST\_CPU} -- number of \code{rtimer} ticks spent in active mode ($2.703\,\mu$A average current consumption on SPES-2 devices).
  \end{itemize}
  The duration of one \code{rtimer} on SPHERE platforms is 1/65536 seconds. Normally, the rest of values can be ignored, as in SPHERE they are insignificant compared to the CPU and radio energy consumption.

\linesep
Packet count message. This should be used in conjunction with Contiki energest status message to infer the total energy consumption of the node.

\begin{lstlisting}[language=json]
{
	'_id': ObjectId('5a608eb9b190070a60fdab2e'),
	'uid': 'fd00::212:4b00:0:ff80',
	'bt': datetime.datetime(2018, 1, 18, 12, 10, 33, 177000),
	'tso': 0,
	'e': [{
		'n': 'PACKETS_TX_UNICAST',
		'v': [0, 0, 0, 0, 0, 0, 0, 0, 0, 0, 0, 0, 0, 0, 0, 0, 0, 0, 354, 0, 3535, 0, 356, 915, 0, 707, 0, 4725, 347, 231, 5183, 13609]
	}, {
		'n': 'PACKETS_TX_BROADCAST',
		'v': [0, 0, 0, 0, 0, 0, 0, 0, 0, 0, 0, 0, 0, 0, 0, 0, 0, 0, 0, 0, 0, 0, 0, 0, 0, 0, 0, 0, 0, 0, 0, 0]
	}],
	'ts': 151627743317.7867,
	'hid': '0000',
	'Date': datetime.datetime(2018, 1, 18, 12, 10, 33, 188000)
}
\end{lstlisting}
  \begin{itemize}
  \item \code{PACKETS\_TX\_UNICAST} -- number of unicast packets transmitted.
  \item \code{PACKETS\_TX\_BROADCAST} -- number of broadcast packets transmitted.
  \item \code{PACKETS\_RX\_UNICAST} -- number of unicast packets received (and normally acknowledged).
  \item \code{PACKETS\_RX\_BROADCAST} -- number of broadcast packets received.
  \end{itemize}
The number of packets is given as an array. Each entry in that array counts number of packets with a specific size: i.e. the first entry counts packets that are 0--3 bytes long, the second for packets 4--7 bytes long, and so on.

\linesep
RPL status message:

\begin{lstlisting}[language=json]
{
	'_id': ObjectId('5a608ef0b190070a60fdadd0'),
	'uid': 'fd00::212:4b00:0:ff81',
	'bt': datetime.datetime(2018, 1, 18, 12, 11, 28, 623000),
	'tso': 0,
	'e': [{
		'n': 'RPL_ROOT_REPAIRS',
		'v': 0
	}, {
		'n': 'RPL_LOCAL_REPAIRS',
		'v': 2
	}, {
		'n': 'RPL_LOOP_ERRORS',
		'v': 0
	}, {
		'n': 'RPL_RESETS',
		'v': 0
	}, {
		'n': 'RPL_LOOP_WARNINGS',
		'v': 0
	}, {
		'n': 'RPL_PARENT_SWITCH',
		'v': 7
	}, {
		'n': 'RPL_MEM_OVERFLOWS',
		'v': 0
	}, {
		'n': 'RPL_GLOBAL_REPAIRS',
		'v': 0
	}, {
		'n': 'RPL_MALFORMED_MSGS',
		'v': 0
	}, {
		'n': 'RPL_FORWARD_ERRORS',
		'v': 0
	}],
	'ts': 151627748862.37112,
	'hid': '0000',
	'Date': datetime.datetime(2018, 1, 18, 12, 11, 28, 625000)
}
\end{lstlisting}
This contains various RPL statistics, for more information consult Contiki 3.x documentation and source code.

\linesep
IP status message:

\begin{lstlisting}[language=json]
{
	'_id': ObjectId('5a60b979b190070a6000475e'),
	'uid': 'fd00::212:4b00:0:ff80',
	'bt': datetime.datetime(2018, 1, 18, 15, 12, 57, 953000),
	'tso': 0,
	'e': [{
		'n': 'IP_FWD',
		'v': 0
	}, {
		'n': 'IP_SENT',
		'v': 28621
	}, {
		'n': 'IP_CHKERR',
		'v': 0
	}, {
		'n': 'IP_RECV',
		'v': 24013
	}, {
		'n': 'IP_VHLERR',
		'v': 0
	}, {
		'n': 'IP_FRAGERR',
		'v': 0
	}, {
		'n': 'IP_PROTOERR',
		'v': 0
	}, {
		'n': 'IP_LBLENERR',
		'v': 0
	}, {
		'n': 'IP_HBLENERR',
		'v': 0
	}, {
		'n': 'IP_DROP',
		'v': 0
	}],
	'ts': 151628837795.3567,
	'hid': '0000',
	'Date': datetime.datetime(2018, 1, 18, 15, 12, 57, 955000)
}
\end{lstlisting}
This contains various IP statistics, for more information consult Contiki 3.x documentation and source code.

\linesep
Link-layer status message:

\begin{lstlisting}[language=json]
{
	'_id': ObjectId('5a60b979b190070a6000475e'),
	'uid': 'fd00::212:4b00:0:ff80',
	'bt': datetime.datetime(2018, 1, 18, 15, 12, 57, 953000),
	'tso': 0,
	'e': [{
		'n': 'LL_TOOLONG',
		'v': 0
	}, {
		'n': 'LL_TOOSHORT',
		'v': 0
	}, {
		'n': 'LL_BADSYNCH',
		'v': 0
	}, {
		'n': 'LL_BADCRC',
		'v': 0
	}, {
		'n': 'LL_CONTENTIONDROP',
		'v': 0
	}, {
		'n': 'LL_TX',
		'v': 2610
	}, {
		'n': 'LL_RX',
		'v': 1259
	}],
	'ts': 151628837795.3567,
	'hid': '0000',
	'Date': datetime.datetime(2018, 1, 18, 15, 12, 57, 955000)
} 
\end{lstlisting}
This contains various link-layer statistics, for more information consult Contiki~3.x documentation and source code.

\subsection{Alert messages}

Gateway script start/stop message example:

\begin{lstlisting}[language=json]
{
	'_id': ObjectId('59d74895e1bce509f7ecf00b'),
	'bt': datetime.datetime(2017, 10, 6, 9, 10, 45, 493000),
	'uid': 'gateway_script',
	'type': 'ALERT',
	'e': [{
		'v': 'START',
		'n': 'ACTION'
	}],
	'hid': '0000',
	'Date': datetime.datetime(2017, 10, 6, 9, 10, 45, 495000)
}
\end{lstlisting}

  \begin{itemize}
  \item \code{ACTION} -- either \code{START} or \code{STOP}, signals that the gateway script has been started or stopped.
  \item \code{TIMESTAMP\_RESYNC\_TIME} -- the last time when timestamps were synchronized between the gateway script and the TSCH network. The value 0.0 means ``never, since the start of the script''. (This field not present in the example above.)
  \end{itemize}

\subsection{Calibration messages}

Water sensor calibration data:

\begin{lstlisting}[language=json]
{
	'_id': ObjectId('59d769067ea12b738b400908'),
	'src': 'piezo_calibrator',
	'jsonrpc': '2.0',
	'dest': 'gateway_script',
	'id': '79388',
	'bt': '2017-10-06T12:29:05.782Z',
	'params': {
		'PIEZO_HOT_END': [2190, 2066, 2106, 2135, 2104, 2118, 2188, 2160, 2162, 2184, 2164, 2107, 2173, 2141, 2170, 2119, 2138, 2124, 1995, 1903, 1912, 1904, 1906, 1899, 1923, 1938, 1938, 1952, 1933, 1931, 1936, 1922, 1937, 1901, 1903, 2013, 2035, 2010, 2014, 1998, 1982, 1998, 2005, 2026, 1839, 1968, 1972, 1960, 1985, 1955, 1966, 1962, 2016, 2029, 2009, 2036, 2038, 2029, 1987, 2006, 2016, 1983, 1991, 1993, 1976, 1640, 1637],
		'PIEZO_INTERRUPT_PIN': [0, 0, 0, 0, 0, 0, 0, 0, 0, 0, 0, 0, 0, 0, 0, 0, 0, 0, 0, 1, 1, 1, 1, 1, 1, 1, 1, 1, 1, 1, 1, 1, 1, 1, 1, 0, 1, 0, 1, 1, 0, 1, 1, 1, 1, 1, 1, 1, 1, 1, 1, 1, 1, 1, 1, 1, 0, 1, 1, 1, 1, 1, 1, 1, 1, 2, 2],
		'PIEZO_HOT_START': [1864, 1858, 1857, 1861, 1858, 1858, 1859, 1861, 1862, 1860, 1856, 1860, 1861, 1863, 1856, 1862, 1858, 1859, 1854, 1763, 1739, 1748, 1738, 1745, 1774, 1794, 1798, 1852, 1745, 1797, 1818, 1797, 1804, 1784, 1761, 1862, 1823, 1858, 1842, 1850, 1863, 1813, 1836, 1839, 1745, 1808, 1809, 1818, 1820, 1786, 1840, 1768, 1858, 1849, 1825, 1802, 1861, 1844, 1828, 1829, 1800, 1794, 1848, 1826, 1799, 1637, 1638],
		'PIEZO_GROUND_TRUTH': [1, 1, 1, 1, 1, 1, 1, 1, 1, 1, 1, 1, 1, 1, 1, 1, 1, 0, 0, 0, 0, 0, 0, 0, 0, 0, 0, 0, 0, 0, 0, 0, 0, 0, 0, 0, 0, 0, 0, 0, 0, 0, 0, 0, 0, 0, 0, 2, 2, 2, 2, 2, 2, 2, 2, 2, 2, 2, 2, 2, 2, 2, 2, 2, 2, 3, 3],
		'PIEZO_COLD_START': [1810, 1699, 1717, 1737, 1736, 1752, 1749, 1734, 1735, 1744, 1716, 1737, 1699, 1709, 1717, 1712, 1732, 1726, 1696, 1847, 1856, 1862, 1856, 1854, 1854, 1853, 1840, 1852, 1856, 1859, 1836, 1856, 1855, 1851, 1862, 1849, 1853, 1743, 1851, 1856, 1840, 1851, 1856, 1858, 1860, 1862, 1854, 1854, 1861, 1853, 1852, 1851, 1856, 1854, 1853, 1856, 1818, 1850, 1855, 1850, 1851, 1854, 1854, 1853, 1853, 1639, 1639],
		'PIEZO_COLD_END': [1938, 1875, 1835, 1855, 1860, 1870, 1883, 1884, 1909, 1830, 1873, 1886, 1871, 1867, 1867, 1866, 1890, 1828, 1804, 2096, 2081, 2110, 2043, 2066, 2125, 2033, 2033, 2058, 2057, 2072, 2104, 2140, 2101, 2122, 2101, 1986, 1965, 2062, 2042, 2039, 1961, 2038, 1980, 1987, 1977, 2055, 2059, 2091, 2073, 2065, 2029, 2077, 2049, 2066, 2079, 2029, 2050, 1986, 2013, 2120, 1997, 2037, 2024, 2022, 1984, 1639, 1638]
	},
	'Date': '2017-10-06T11:29:05.124Z',
	'hid': '0000',
	'method': 'piezo_calibration_set'
}
\end{lstlisting}

  \begin{itemize}
  \item \code{PIEZO\_COLD\_START} -- cold flow start values.
 \item \code{PIEZO\_COLD\_END} -- cold flow end values.
 \item \code{PIEZO\_HOT\_START} -- hot flow start values.
 \item \code{PIEZO\_HOT\_END} -- hot flow end values.
 \item \code{PIEZO\_INTERRUPT\_PIN} -- interrupt status values.
 \item \code{PIEZO\_GROUND\_TRUTH} -- ``ground truth'' class labels marked by the technician who calibrated the sensor upon installation.
   \end{itemize}

\subsection{Other data}

SPHERE also uses control messages and other ephemeral MQTT messages to exchange commands and internal data between the system components. However, these messages are not part of the dataset.

%% file: dataquality.tex
\section{Data quality issues}
\label{sec:dq}

\subsection{Missing data}

Data can be missing because of multiple reasons:

\begin{itemize}
\item System switched off or paused. If just paused, monitoring data is still going to be present.
\item Home Gateway NUC broken or otherwise not recording data in the database. Monitoring data still may be present.
\item A subsystem is not producing data. E.g. the gateway script may be stopped, which means that there is going to be neither wearable, nor environmental sensor, nor network monitoring data.
\item A single device is not producing data. E.g. a video NUC switched off or rebooting; video camera switched off; an environmental sensor has run out of battery; a gateway node is not powered; a wearable is out of the house.
\item A component of a single device is not working. E.g. some of the gateways occasionally stop forwarding wearable data, even though they still produce and forward environmental sensor data.
\end{itemize}
In a single house, all environmental sensors have somewhat similar battery lifetime. E.g. if a house loses the first environmental sensor after 6 month operation, it's likely that by the end of the month several more will follow, and after 2 more months either all sensors are going to be out of battery, or their batteries will have been replaced by SPHERE technicians.

\subsection{Transient failures}

The SPES-2 and SPG-2 sensor nodes occasionally reboot. Compared to the initial state of the SPHERE hardware platforms that were rebooting many times per day, by the time of the first public deployments this number of reboots is reduced by several orders of magnitude. Still, there is on average approximately one reboot per node per month.

\begin{itemize}
\item Reboot of the main gateway node. This one has the most impact, as data from all of the SPES-2, SPG-2 and SPW-2 nodes is lost for a while. It takes 3 minutes for the other nodes to notice that the old network is gone, and try to start joining the new one. The joining itself may take several minutes; the actual time depends on many random factors and can be very different for different nodes.
\item Reboot of another SPG-2 node. The data from this node (including SPW-2 data that it is supposed to be forwarding) is lost until the node rejoins the network, which can take up to several minutes.
\item Reboot of SPES-2 node similarly means it does not produce sensor data until rejoining the network. Since SPES-2 nodes are battery powered, and the join process takes a lot of energy, frequent reboots may affect SPES-2 battery life.
\end{itemize}

\subsection{Timestamps}

\subsubsection{General notes}

In general, it is tricky to produce timestamps that are always correct. A fault in any of these steps (even a transient fault) is going to cause timestamps that are invalid for a while. E.g. when the root gateway node reboots and the time counter of the TSCH network resets, there might be few documents with invalid \texttt{bt} values, corresponding to the old time offset before the restart of the TSCH network.

A good practice when handling SPHERE data is to compare the \texttt{bt} value (created at the data source) with the \texttt{Date} value (if present -- it is only present in monitoring and control data) and the \texttt{ObjectID} field (created by the MongoDB). The \texttt{Date} field describes the time the data was received by the MQTT broker. From the \texttt{ObjectID} field, the database record creation time can be recovered.
If there is a more than a few minutes mismatch for a document, discard that document. Example code:

\begin{lstlisting}
import calendar
import time

MAX_BT_DIFF = 10 * 60 # the maximal allowed difference of timestamps

# `doc` is a SPHERE JSON document
bt = doc['bt']
objid = doc['_id'].generation_time
# convert the value to a GMT timestamp - all SPHERE documents use GMT/UTC timestamps
bt_ts = calendar.timegm(bt.timetuple())
# convert the `objid` value to a timestamp
objid_ts = time.mktime(objid.timetuple())

delta = abs(bt_ts - objid_ts)
if delta > MAX_BT_DIFF:
   print('Too big time difference, discarding the document...')
\end{lstlisting}
From analyzing a subset of the data, we estimate that approximately 0.08\,\% of documents have mismatched timestamps between the sensor network and MQTT time (the \texttt{bt} and \texttt{Date} values), and 0.25\,\% have mismatched timestamps between MQTT and the Mongo database values (the \texttt{Date} and \texttt{ObjectID} values).

\subsubsection{Time synchronization architecture}

The time synchronization architecture in SPHERE is as follows:
\begin{itemize}
\item The SPHERE Home Gateway is NTP synced to the University of Bristol time server.
\item The other SPHERE components are synced to the SPHERE Home Gateway (the SHG is an NTP server for video NUC).
\item The TSCH network has its own time counter. This counter is reset every time the TSCH root gateway node restarts. (This can happen as often as multiple times per month.) This counter is not synchronized with the SPHERE Home Gateway directly, but an offset is added to the documents during their processing on the SPHERE Home Gateway. This offset describes the time difference between the TSCH network and the UTC time.
\end{itemize}

\subsubsection{Timestamps for environmental sensor data}

Timestamps for the environmental sensor data are with resolution of  $0.01$ seconds. However, since environmental values change slowly, the timestamp accuracy for these readings in SPHERE is not high. For data compression reasons multiple environmental sensor readings can reuse the same timestamp, as long as the difference between the reading is less or equal to 1~second. (Note that this also applies to PIR sensor and water flow sensor readings.)

\subsubsection{Timestamps for wearable data}

The wearable data are time-stamped at the SPHERE Forwarder nodes. The accuracy of the timestamps are bounded by three types of timing errors (the propagation and processing delays are insignificant, \textit{i.e.} less than $1$~ms).
The first source of timing errors originates from the fact that, in the TI RTOS BLE stack, advertisements are scheduled using a different clock source than the primary application processor.
As a result, the packet generation and the packet transmissions are scheduled with different clocks that drift.
This leads to a delay that is bound by the period of the advertisements which in turn depends on the sampling frequency (\textit{i.e.}, $0.48$ seconds at $12.5$~Hz and $0.24$ seconds at $25$~Hz). 
The second source of timing errors is due to queuing delay in the SPHERE Forwarders. This delay is measured at $0.05$ seconds in the worst case scenario.
The last source of the timing errors originates from the TSCH timestamps which have $0.01$ second granularity.

\subsection{Environmental sensor data}

\begin{itemize}

\item BMP and HDC sensors may stop working or occasionally produce invalid readings. This will be seen in the dataset as absence of data from that particular sensor or invalid data. One such case is that the HDC sensor on some nodes reads 0 instead of the expected raw data value. When this raw value 0 is converted to \degc temperature for storing in the JSON document, it ends up being equal stored in the database as $-$40~\degc.

\item Light sensors and PIRs may become dusty, dirty or get covered by the participant, and stop producing valid readings or change their dynamic range because of there reasons.

\item Light sensors may be exposed to direct sunlight or direct light from indoor data sources. While not a fault as such, this may create unexpected data, as the sensor is supposed to record ambient light levels. This will not normally happen since the light sensor is \emph{inside} the enclosure; it may occasionally happen since its close to the openings in the sensor enclosure.

\item PIRs may be facing windows; in this case movements of clouds are going to trigger the PIR. Although the technicians are instructed not to place PIRs opposite windows, in some cases it might be unavoidable.

\item Temperature sensors of the SPG-2 gateway nodes always report 1--2~\degc higher temperature than those of SPES-2 nodes. The  mechanism of how this happens is not fully clear. One workaround: use only relative temperature values.

\item Temperature sensors may be placed near heat source and record their data instead of the ambient temperature data. Again, while this is not a fault as such, it may create unexpected sensor readings.

\item The water sensor may be incorrectly calibrated. Some of the SPHERE deployments faced calibration problems. For those installations, the classification results will be missing, but the raw water sensor data is going to be recorded in the database anyway.

\item The water sensor clamp attachments to water pipes may change over time (\eg become loose), rendering the calibration data unsuitable for results after the change.

\end{itemize}
One way to work around invalid sensor readings is sanity-checking whether the data falls in reasonable bounds. (We do not discard these ``invalid'' readings as they also could signal that a real, exceptional event has happened in the house.) For example, these values we are using internally:

\begin{eqnarray*}
0 \text{~\degc}  \leq & \text{HDC\_TEMP} & \leq 50 \text{~\degc} \\
0 \text{~\%}  \leq & \text{HDC\_HUM} & \leq 100 \text{~\%} \\
0 \text{~\degc}  \leq & \text{BMP\_TEMP} & \leq 50  \text{~\degc} \\
950 \text{~hPa} \leq & \text{BMP\_PRES} & \leq 1050 \text{~hPa} \\
0 \text{~lux}   \leq & \text{LT} &  \leq 10000 \text{~lux} \\
\end{eqnarray*}

On top of that, median filter can be applied to discard occasional noisy readings. However, we haven't detected a significant amount of random errors in the environmental sensor data (unlike the wearable data).

\subsection{Wearable acceleration data}
\label{sec:data-quality-wearable-accel}

Some of the data gathered from wearables have random errors. Since these error usually affect only single reading, these values can be filtered out with a median filter. Note that this problem is not limited to acceleration data; other data created on wearables are  affected by this as well, e.g. message sequence numbers and field in monitoring data, such as battery voltage.

Let's assume the data has been read into a Python iterable. Then one way to apply the median filter is:

\begin{lstlisting}
def median_filter(data):
    result = [0] * len(data)
    result[0] = data[0]
    for i in range(1, len(data) - 1):
       result[i] = median(data[i-1], data[i], data[i+1])
    result[-1] = data[-1]
    return result
\end{lstlisting}

    The are multiple ways to implement the \code{median()} function. On option is to use sorting: \code{return sorted([a, b, c])[1]}. However, sorting the data is not the most optimal way to compute the median. For three values, this series of explicit \code{if/else} statements is a faster and more portable method:

\begin{lstlisting}
def median(a, b, c):
    if a > b:
        if b > c:
            return b # a, b, c
        if a > c:
            return c # a, c, b
        return a # c, a, b
    else: # a <= b
        if a > c:
            return a # b, a, c
        if b > c:
            return c # b, c, a
        return b # c, b, a
\end{lstlisting}

\begin{figure}%
\centering
\subfloat[A 20 minute segment during a sleeping activity where the received acceleration has been corrupted (approx at 00:14).]{\includegraphics[width=0.45\linewidth]{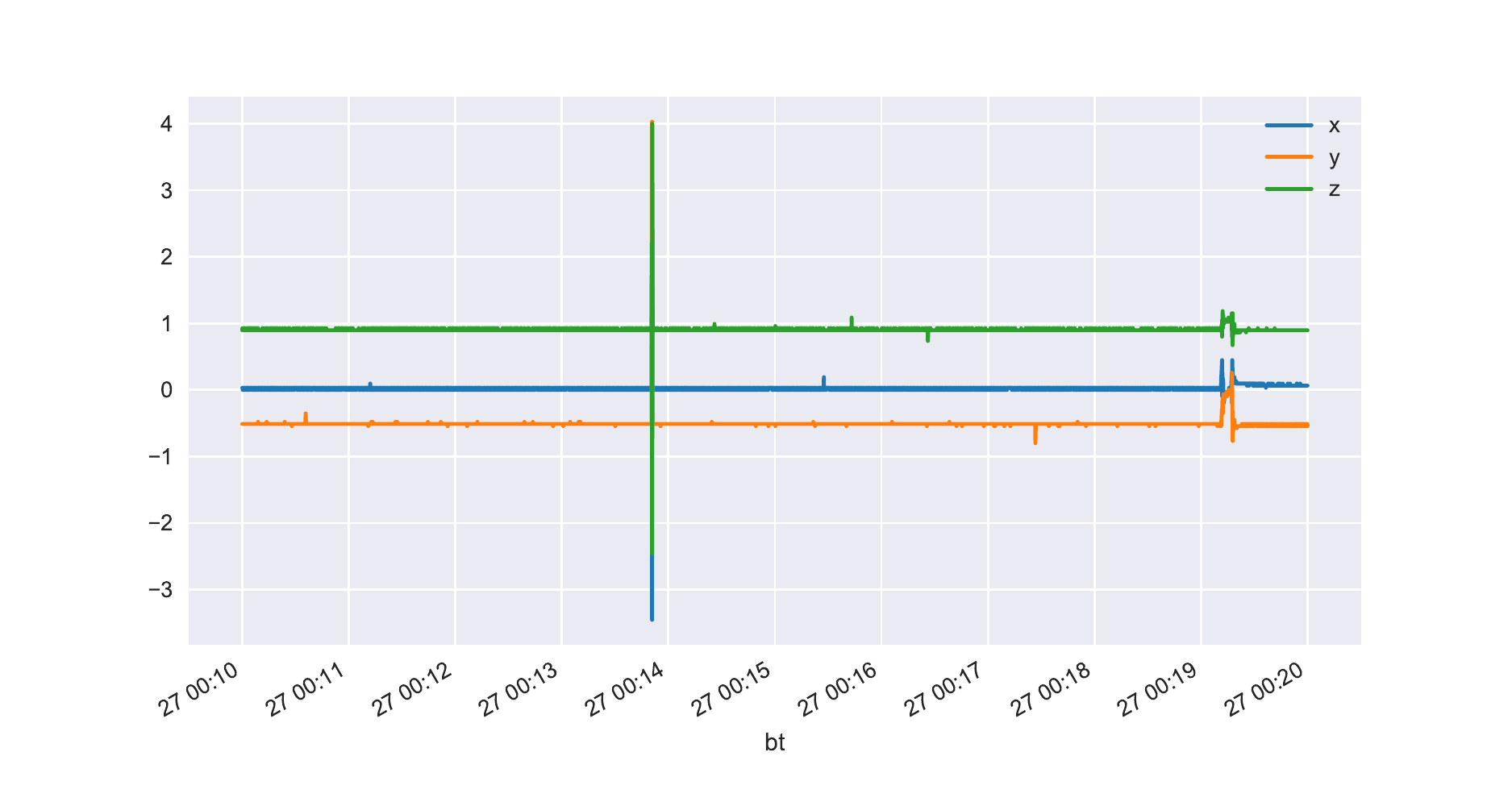}\label{fig:zoomed_out_accel}} ~
\subfloat[A 5 second segment illustrating the effect of the accelerometer corruption on the received signal. Notice the corruption begins after 00:13:50 and persists for 6 samples.]{\includegraphics[width=0.45\linewidth]{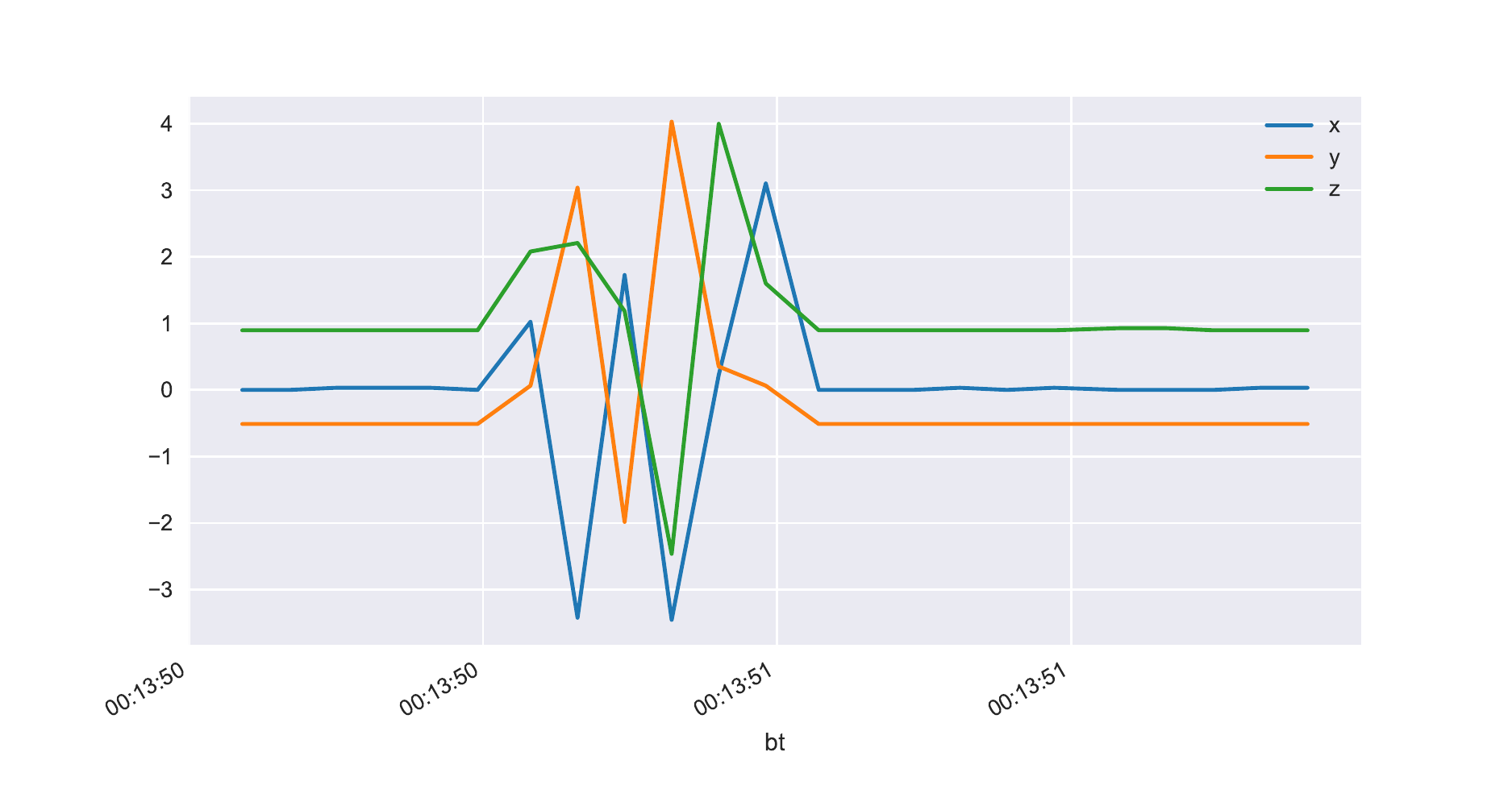}\label{fig:zoomed_in_accel}}%
\caption{This figure illustrates the effect of corruptions on the acceleration data.}
\label{3figs}
\end{figure}

Of interest also are circumstances where the median filter is unable to recover from these issues. Although these are rare events they predominantly occur when packets arrive with low RSS. Figure \ref{fig:zoomed_out_accel} shows an example of the corrupted data. This was taken from a period during which the participant was sleeping, and hence we would expect to see data acceleration data to remain relatively steady. However, at approximately 00:14 we can notice an abrupt disturbance to the signal. We focus only on this segment in Figure \ref{fig:zoomed_in_accel} and can count that this disturbance persists for 6 samples. We can conclude then from this these corruptions arise from one received packet since these contain six samples of accelerometer data. 

Since these disturbances are of a large magnitude they may affect the feature representations that are necessary for analytics, activity of daily living detection, behavioural monitoring \textit{etc}. Several techniques may be applied in combatting the effect here. First one can consider filtering these signals with a low-pass filter since these are predominantly made up of high-frequency data outside of the range of human movement. One may also aggregate the data at the packet level and perform anomaly detection on the sequence of received packets and reject data which is considered anomalous. With the above example in particular, but also with acceleration traces resulting from normal activity within the home, the characteristics of the corrupted data are sufficiently large that any standard anomaly detection techniques will identify the samples as irregular.

\subsection{Wearable localization data (RSSI)}

\subsubsection{SPG-2 recordings}

On occasion, a selection of the SPG-2 nodes do not receive any wearable advertisement packets and require a manual reboot to resume normal operation.
Nodes may be found in the SPHERE dataset which do not have advertisement data for a significant portion of the deployment duration, even when wearables were present in the house. 
Non-reception of wearable advertisement packets does not mean that there is a fault with the hardware as the wearable can often be out of communication range of a single SPG-2 node.
If processed data from the other nodes and other sensors indicated that the wearable was co-located in the same room as the non-functioning node, a hardware fault is the most likely explanation.


\subsubsection{NUC recordings}
\label{sec:data-quality-nuc-recording}

The BLE reception quality by the NUCs is significantly poorer than that of the SPG gateway nodes. Wearable packets are often only received at much shorter ranges and lower rates than the custom SPG receivers. NUC BLE antennas are embedded within the device, shielded by other components, and designed for communication with peripherals located within a few meters of the device. As a result, in some Homes 
there are few wearable recordings registered on the NUCs.

\begin{figure}
  \centering
  \includegraphics[width=0.7\linewidth]{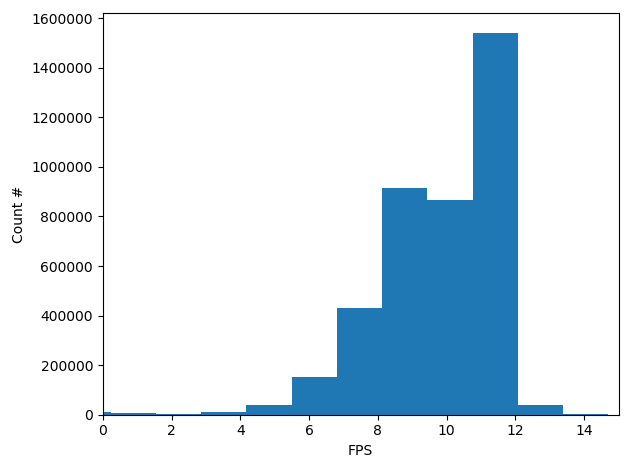}
  \caption{Histogram of frame rate for the video recorder.} 
  \label{fig:fps}
\end{figure}

The data may also get lost on the WiFi connection between the Video NUC and the SPHERE Home Gateway.
One of the causes is the variability on the frame rate (see Figure~\ref{fig:fps}), which are usually caused by the camera sensors' drivers. Together with other challenging environmental causes, this may affect to the long-term transference of packages.
To calibrate the cameras with respect to the changing ambient lighting conditions throughout the day, the NUCs are rebooted at three different times each day: 06:30, 10:30 and 00:00.

A NUC only picks up RSSI data, not acceleration data, as they do not have the decryption keys for acceleration data.
If a packet has been received only by a NUC and not by any of the gateway nodes, a record is added to the database without the acceleration data (\ie the \code{e} array in the document is empty, only \code{gw} is filled).

\subsubsection{Pitfalls in interpreting RSSI values}

The relation between RSSI and location is not straightforward. Indoor environments are usually modeled as multipath RF propagation environments, with shadowing, reflections and other non-line-of-sight propagation phenomena taking place.

The following factors affect signal strength; they are ordered by their approximate magnitude: 
\begin{itemize}
\item The direct distance between the wearable and the receiver, especially in line-of-sight conditions.
\item Environmental shadowing and reflections. Walls, ceilings and large pieces of furniture will significantly alter and inhibit signal propagation. The material of the obstacles has a direct influence on whether the signal is attenuated or diffracted: e.g. metal causes reflection or diffraction, the signal is attenuated significantly through brick or stone, but less so through wood. The building materials of the \emph{SPHERE 100 Homes Study} houses was captured by the site-survey visits and we expect that metadata to be available for the users of the SPHERE dataset(s).
\item Body shadowing. Signal strength is significantly affected when the wearable is directed away from the receiving node and the wearer's body blocks the communication path, for example, a participant standing in the same location and rotating 360$^\circ$ would register RSSI readings that alter significantly, resembling  a sine wave curve when plotted.
\item Antenna polarization. Changing the angle the wearable is held relative to the receiver's antenna position affects the RSSI. For example, rotating the hand with a wearable around it's axis is going to significantly affect RSSI even if the rest of the body remains still. 
\item Fading, such as deep fading caused by multipath propagation of the signal. 
As this type of fading is frequency-selective, aggregating the signal over multiple frequencies (e.g. with \texttt{mean()} or \texttt{max()} from multiple samples should help to mitigate this issue.
Note that the SPHERE dataset does \emph{not} include information on which frequency a particular message was received along with the RSSI value of that message due to an unfortunate limitation of the Texas Instruments RTOS (Real Time Operating System) BLE (Bluetooth Low Energy) stack. To overcome this, a time window can be selected and appropriate aggregates calculated for that time bin. A large enough time window will ensures that RSSI values in \emph{all} of the three advertisement frequencies are considered, however there is a trade-off when considering the time resolution of the participant's motion or activities.
\item A dynamic environment, e.g. closing or opening doors, moving furniture and other household items all affect the signal strength through changes in signal reflection/propagation environment.
\item Random drift over time, possibly caused by the hardware itself. In terms of distance to the transmitter, this noise behaves as a constant, meaning that it is more likely to be caused by the receiver, not the transmitter. It also means that it affects weak RSSI values much more strongly; i.e. the effect from this rarely registers on $-50\,$dBm values RSSI, but can be prevalent on $-80\,$dBm RSSI.
\item The way the wearable is worn. Wearing it on a tighter strap puts in closer to skin, can change the assumed propagation characteristics of the wearable antenna when in closer proximity to a significant dielectric like the body.
\item External interference from other Bluetooth and other 2.4GHz devices. Early lab tests do not fully indicate the full effect of unwanted transmitting sources, however, from initial tests, there appears to be a slight increase of the standard deviation of the signal. 
\end{itemize}

In contrast, there is no reason to believe that these factors have significant effect on RSSI: 
\begin{itemize}
\item Battery voltage levels on wearables and other battery-powered devices. The voltage of the radio subsystem is regulated by the CC2650 chip, and is kept stable even when the battery voltage goes down.
\item Environmental temperature and humidity. The SPHERE networks are deployed in room conditions, with stable $20-30~^o$C temperature. The humidity levels, in contrast, can be quite variable indoors, but the SPHERE wireless links are too short for this to have significant effect.
\end{itemize}
In theory, other causes of RSSI variation may be present, such as the receiver locking in on a specific channel. In this case it will receive single-frequency signals from the transmitter rather than hopping over all available frequencies pseudorandomly.
In this case, the aggregate-over-time-window approach would not help, as an impractically long time window would have to be selected for calculating the aggregate RSSI. However, the authors believe that this did not occur in the SPHERE dataset. 

\subsection{Video data}

Despite some quality issues described in section~\ref{sec:data-quality-nuc-recording} referred to missing packages, the approaches for extracting the silhouettes and visual features from RGB and Depth are suitable for lightweight deployment. However, due to the high variability of conditions captured by these sensor data, there are situations where the quality of visual features and silhouettes are far from accurate. Therefore, this limitation has to be specially taken into account when relying on these features for posterior analyses.

\subsection{Monitoring data}

\begin{itemize}
\item Wearable: see Section~\ref{sec:data-quality-wearable-accel}.


\item Some data is published only to \texttt{MON} topics, therefore it is not saved in the \texttt{RAW\_MON} topics in the database along with most of the other monitoring data. It is still going to be part of the data received over the 3G connection. One example of such data is  notification message data published by BLE scripts (the scripts that pick up wearable advertisements on NUCs).

\item Version E adds some new monitoring data fields that are missing in version D, for example, the \texttt{GW\_NODE\_ALIVE} field in the gateway script's heartbeat messages.
\end{itemize}
